\documentclass{article}

\usepackage{microtype}
\usepackage{graphicx}
\usepackage{subcaption}
\usepackage{booktabs} 
\usepackage{enumitem}

\usepackage{hyperref}

\usepackage[accepted]{icml2026}

\usepackage{amsmath}
\usepackage{amssymb}
\usepackage{mathtools}
\usepackage{amsthm}

\usepackage[capitalize,noabbrev]{cleveref}

\theoremstyle{plain}

\theoremstyle{definition}

\theoremstyle{remark}

\usepackage[textsize=tiny]{todonotes}

\icmltitlerunning{Spectral Diffusion}

\begin{document}

\twocolumn[
  \icmltitle{Spectral Diffusion for Protein Dynamics}

  \icmlsetsymbol{equal}{*}

  \begin{icmlauthorlist}
    \icmlauthor{Hew Phipps}{stats}
    \icmlauthor{Matteo Cagiada}{stats,denm}
    \icmlauthor{Santiago D. Villalba}{bayer}
    \icmlauthor{Charlotte M. Deane}{stats}
  \end{icmlauthorlist}

  \icmlaffiliation{stats}{Department of Statistics, University of Oxford, Oxford, United Kingdom}
  \icmlaffiliation{bayer}{Bayer, Berlin, Germany}
  \icmlaffiliation{denm}{Department of Biology, Copenhagen University}

  \icmlcorrespondingauthor{Charlotte Deane}{charlotte.deane@stats.ox.ac.uk}

  \icmlkeywords{Machine Learning, ICML, diffusion, dynamics, molecular dynamics, protein, structure, Fourier transform, discrete cosine transform, spectral, deep learning, emulator}

  \vskip 0.3in
]

\printAffiliationsAndNotice{}  

\begin{abstract}
      Generative models present a promising alternative to expensive molecular dynamics for computationally querying protein dynamics, yet many existing approaches treat ensembles as unordered snapshots rather than temporally coherent trajectories, or scale poorly with protein size. We present a new physics-informed representation using Fourier transforms as an inductive bias for the multiscale temporal nature of protein dynamics. Diffusion in the spectral domain allows for disentangling of dynamics into slow conformational modes and fast atomic jitter, enabling rapid and improved prediction of dynamics across a range of temperatures. This is facilitated by denoising of structure and temperature conditioned spectral volumes where the low frequencies directly encode per-residue flexibility. Trained on the mdCATH dataset, we evaluate our model, DynaMode, on a held-out test set achieving strong performance across a set of ensemble-based metrics including a Root Mean Squared Fluctuation (RMSF) pearson $r$ of $0.844$. Code is available at \url{https://github.com/HPuntu/DynaMode}. 
\end{abstract}

\begin{figure*}[!t]
  \centering
  \includegraphics[width=\textwidth]{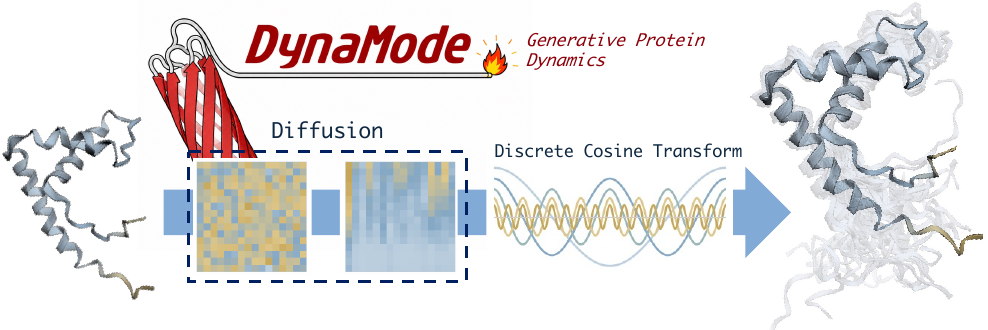}
  \caption{DynaMode is a diffusion model that iteratively denoises a spectral volume representation of protein dynamics given an input structure and temperature. The predicted spectral volume is inverse DCT-II transformed into a trajectory of structures over time.}
  \label{fig:abstract}
\end{figure*}

\section{Introduction} 
Proteins can dynamically adopt diverse conformational states that are often difficult to capture at high resolution with experimental methods. Where experimental approaches are limited and difficult to interpret, computationally modeling protein dynamics can offer insights into key biological processes such as folding, binding and allostery~\cite{huynhProteinDynamicsBased2025,noeBoltzmannGeneratorsSampling2019c}. Traditionally, Molecular Dynamics (MD) numerically solves the Newtonian mechanics of a protein structure over time but is computationally expensive. Enhanced sampling methods have been used to accelerate conformational space exploration but their reliance on system specific collective variables makes them less general~\cite{heninEnhancedSamplingMethods2022,zhuEnhancedSamplingAge2026}. Other modeling approaches, such as G\~{o}-based Ising-like models, use state-based discretisations based on binary native-like contact formation, but these are often poor approximations of the underlying N-body chain mechanics~\cite{takadaGoModelRevisited2019,jiangModelingStructuralFlexibility2012}.

With the success of generative models in structure prediction from sequence~\cite{jumperHighlyAccurateProtein2021e}, methods like diffusion and flow-matching offer a cheap alternative to solving dynamics numerically by learning to instead predict MD trajectories directly from structure using deep neural networks~\cite{lewisScalableEmulationProtein2025b,jansonDeepGenerativeModeling2025}. Such "MD emulators" hold potential for a range of possible applications across temporal upsampling, interpolation between conformations, and inpainting~\cite{jingGenerativeModelingMolecular2024b}. However, dynamics generation requires sampling long, temporally consistent trajectories where motion is highly autocorrelated and influenced by environmental conditions such as temperature, making it a distinct challenge from structure prediction. 

We sought to address this problem through a change of representation from the time to the spectral domain which has shown success in generative image dynamics~\cite{liGenerativeImageDynamics2024}. The spectral domain provides a powerful inductive bias for temporal dynamics by disentangling multiscale temporal correlations and coupled motions into an orthogonal frequency basis, yielding a better conditioned learning problem~\cite{ngueabouIntegratingSpectralMethods2025}. Additionally, extending this approach to proteins offers unique opportunities for direct per-residue flexibility prediction. 

In this work, we present DynaMode, a diffusion model trained on mdCATH~\cite{mirarchiMdCATHLargeScaleMD2024b} to generate Discrete Cosine Transform (DCT) spectral volumes for monomers under 576 residues at temperatures between 300K to 450K. DynaMode is a general dynamics generator that captures protein motion across temperatures over a set of ensemble metrics whilst maintaining temporal coherence including on out-of-distribution temperature regimes. 

\subsection{Contributions}
\begin{enumerate}[leftmargin=*, itemsep=0pt, topsep=1pt]
    \item \textbf{Fast and Accurate MD Emulation} We achieve superior performance on key metrics for protein dynamics and ensemble properties including an RMSF Pearson of $r=0.844$ on the mdCATH test set and $r=0.734$ on the out-of-distribution ATLAS dataset with sampling times of $\sim1$ seconds per 250 frame trajectory on a GH200 GPU.
    \item \textbf{Spectral Convolution Architecture} We develop a custom spectral convolution architecture inspired by Fourier Neural Operators (FNOs) that enables this rapid sampling speed through block-wise spectral mixing.
    \item \textbf{Per-Residue Flexibility Prediction} We show through Parseval's theorem how the low frequencies analytically approximate RMSF whilst being more expressive measures of residue motion. Through $x_0$ prediction diffusion denoising with an MSE loss the model also functions as a zero-shot RMSF-like per-residue flexibility predictor for a given structure.
    \item \textbf{Spectral Protein Dynamics} We show that the DCT transformation is more robust to discontinuity boundaries in high temperature non-equilibrium protein dynamics than the Discrete Fourier Transform (DFT).
\end{enumerate}

Although our representation provides a powerful inductive bias for dynamics, we note that without a post-inference energy minimisation structural validity suffers.

\section{Background}
\label{sec:background}

\paragraph{Diffusion for Proteins}
\label{sec:diffusion}
Diffusion models provide a general framework for generative models by learning to iteratively reverse a Gaussian corruption process. Let $\mathbf{x}_0 \in \mathbb{R}^d$ denote a generic protein object, such as cartesian $C_\alpha$ coordinates, and let $\mathbf{c}$ denote conditioning information such as sequence, structural context, or simulation conditions. In the DDPM formulation~\cite{hoDenoisingDiffusionProbabilistic2020a} noisy samples are defined as $\mathbf{x}_t = \sqrt{\bar{\alpha}_t}\,\mathbf{x}_0 + \sqrt{1-\bar{\alpha}_t}\,\boldsymbol{\epsilon}$, inducing

\begin{equation}
    q(\mathbf{x}_t \mid \mathbf{x}_0)
    =
    \mathcal{N}\!\left(
        \mathbf{x}_t;
        \sqrt{\bar{\alpha}_t}\,\mathbf{x}_0,
        (1-\bar{\alpha}_t)\mathbf{I}
    \right),
\end{equation}
where $\boldsymbol{\epsilon}\sim\mathcal{N}(\mathbf{0},\mathbf{I})$ and $\bar{\alpha}_t=\prod_{s=1}^{t}(1-\beta_s)$. We adopt the cosine schedule of Improved DDPM~\cite{nicholImprovedDenoisingDiffusion2021}. Rather than predicting the added noise, a denoiser directly predicts the clean object from the noised input,
\begin{equation}
    \hat{\mathbf{x}}_0
    =
    D_\theta(\mathbf{x}_t,t,\mathbf{c}),
\end{equation}

This clean prediction defines the score estimate
\begin{equation}
    s_\theta(\mathbf{x}_t,t,\mathbf{c})
    =
    \frac{
        \sqrt{\bar{\alpha}_t}\,
        D_\theta(\mathbf{x}_t,t,\mathbf{c})-\mathbf{x}_t
    }{1-\bar{\alpha}_t}
    \approx
    \nabla_{\mathbf{x}_t}\log q_t(\mathbf{x}_t\mid\mathbf{c}).
\end{equation}

At inference, we use the deterministic DDIM update~\cite{songDenoisingDiffusionImplicit2022}, permitting a reduced number of reverse steps. Training is performed with a denoising regression objective, and inference proceeds by sampling from high-variance Gaussian noise and applying the learned reverse process. Here, we use the same denoising framework for generative modeling of protein dynamics, but jointly over $T$ structures at once.

\paragraph{Ensemble Samplers}
\label{sec:ensemble_samplers}
Prior work has shown that finetuning existing structure prediction models on MD data improves sampling of multiple conformational states across the MD-derived Boltzmann distribution~\cite{jingAlphaFoldMeetsFlow2024c,lewisScalableEmulationProtein2025b,jansonDeepGenerativeModeling2025}. However, these typically rely on large pretrained structure modules or sequence-based embeddings like ESM that can slow inference. 

Repeated querying of these models to sample an ensemble results in a set of structures without explicit temporal ordering. Such approaches focus on conformational state exploration as opposed to dynamics~\cite{jansonDeepGenerativeModeling2025,jingEigenFoldGenerativeProtein2023a,kapusniakMarSFMGenerativeModeling2026,sengarGenerativeModelingFullAtom2025}. The recent MarS-FM explicitly learns metastable states through defining Markov State Models (MSMs) which guide parallel conformational sampling, resulting in drastically improved sampling speed and conformational space traversal~\cite{kapusniakMarSFMGenerativeModeling2026}. Our approach seeks to explicitly model state sampling over time, which we do so efficiently by sampling whole trajectory windows at once.

\paragraph{MD Emulators}
\label{sec:md_emu}
There has been a recent emergence of so-called MD-emulators which one-shot generate complete trajectories given an input structure, having been effectively demonstrated in small systems over a range of functionalities including upsampling, interpolation and inpainting~\cite{jingGenerativeModelingMolecular2024b}. In our case we are explicitly interested in trajectory generation/extension, which has also been approached by iterative structure sampling as a function of time~\cite{fengBioMDAllatomGenerative2025,xuTEMPOTemporalMultiscale2025b}. These methods tackle long-scale dynamics generation for larger systems with curriculum based sampling, where a "forecaster" or "planner" samples sparsely separated conformations across long timescales, followed by an interpolator.

Rather than representing the multiple scales of protein dynamics by training on different timestep partitions of the data~\cite{xuTEMPOTemporalMultiscale2025b}, here we explicitly expose the slow and fast modes through the DCT transform. Recent success in latent space diffusion models evidences the field is moving in the direction of lower dimensional representations for protein dynamics~\cite{sengarGenerativeModelingFullAtom2025,sengarEnsemblesSimulatingAllAtom2025}. The current state of the art, ATMOS, uses latent states that scale linearly with the number of structures sampled whilst driving temporal sampling for all-atom generative dynamics~\cite{shiAtomicTrajectoryModeling2026}.

\paragraph{Low Dimensional Representations of Protein Dynamics}
\label{sec:low_dim}
MD trajectories are by nature high-dimensional, but proteins exhibit strong spatiotemporal autocorrelations, making it well-established in the biophysics literature to project dynamics onto a small number of collective modes. Methods such as Principal Component Analysis (PCA) and Time-lagged Independent Component Analysis (tICA) identify these slow-mode subspaces directly from simulation data \cite{schultzeTimeLaggedIndependentComponent2021}. Whilst standard tools for MD analysis, their utility for generative modeling is fundamentally limited by their dependence on protein-specific trajectory data to define the basis, precluding generalisation to unseen sequences and motions. 

Separately, Normal Mode Analysis (NMA) recovers collective motions analytically by treating the protein as a system of harmonic oscillators around an energy minimum, providing a simulation-free but equilibrium-specific basis~\cite{bauerNormalModeAnalysis2019}. Analogously, EigenFold formulates protein structure generation around eigenvectors of the residue graph Laplacian, defining a coarse-to-fine diffusion schedule in which low eigenmodes establish global topology before high eigenmodes resolve local geometry~\cite{jingEigenFoldGenerativeProtein2023a}. Latent diffusion on spatial graph Laplacian modes specifically leverages the multiscale nature of protein geometry~\cite{sengarGenerativeModelingFullAtom2025}.

In contrast, Fourier transformation over the time axis offers a universal basis across different proteins and temperatures that is generative, whilst yielding similar multiscale dynamics separation and slow-mode collective motion encoding into the low frequencies. 

\paragraph{Fourier Basis}
\label{sec:fourier}
Spectral representations are standard in mechanics and signal processing, where Fourier transformations offer a natural representation for temporal processes by decomposing trajectories into orthogonal modes ordered by frequency~\cite{canutoSpectralMethodsFundamentals2010,ngueabouIntegratingSpectralMethods2025}. 
Applied along the time axis of protein MD trajectories, the DCT is a real-valued Fourier transform which separates slow, collective conformational changes from faster local fluctuations by assigning them to low- and high-frequency coefficients, respectively~\cite{ahmedDiscreteCosineTransfom1974}. This is consistent with established frequency-domain analyses of MD, including vibrational spectra from Fourier transforms of time-correlation functions and multivariate frequency-domain analysis of protein dynamics~\cite{matsunagaMultivariateFrequencyDomain2009}, as well as trajectory-compression and acceleration using essential-dynamics and PCA or DCT representations~\cite{meyerEssentialDynamicsTool2006,kumarCompressionMolecularSimulation2013,liangAcceleratingMolecularDynamics2026}. 

DCT is particularly suited to finite trajectory windows where it avoids the strong periodicity assumption of DFT which manifests as undesirable spectral leakage of coefficient correlations across the high frequencies~\cite{wallaceJPEGStillPicture1992,yaroslavskyDFTDCTMDCT}. Inspired by recent spectral-volume diffusion models for image dynamics ~\cite{liGenerativeImageDynamics2024}, we use DCT spectral volumes as the predictive target for diffusion.

\begin{figure*}[!t]
  \centering
  \includegraphics[width=\textwidth]{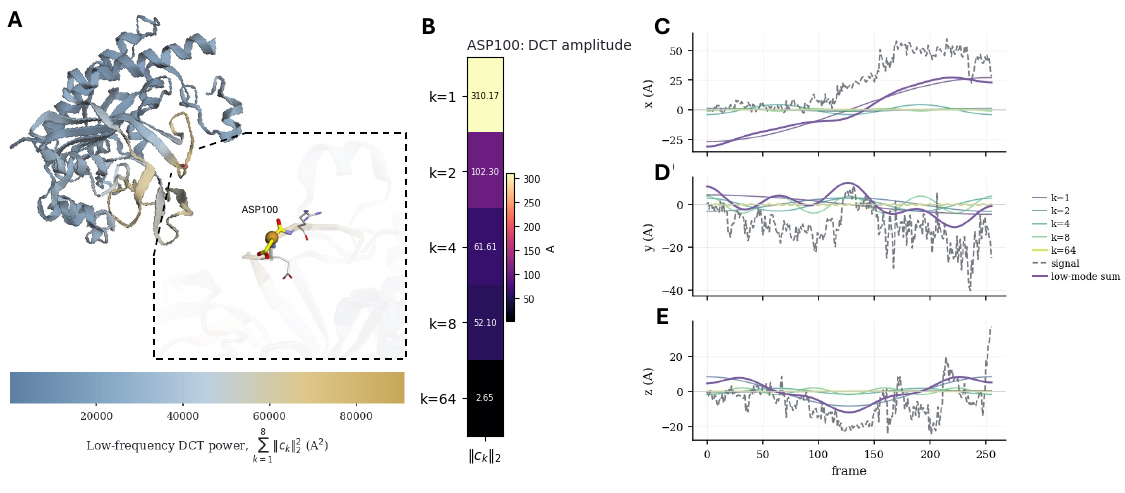}
  \caption{The DCT transform gives $\tau$ frequency coefficients for each $C_\alpha$ coordinate channel (x,y,z) of each residue. \textbf{A} Illustrative cartoon depiction 12asA00, a 327 residue mdCATH domain coloured by the per-residue spectral power of the lowest $k=8$ frequencies over a 450K trajectory. Dashed box: Close up of ASP100, the residue with the highest spectral power. \textbf{B} DCT spectral amplitudes for a number of different frequencies for ASP100. Each frequency is represented by its cosine wave over time  for the x, y and z channels (\textbf{C}, \textbf{D}, \textbf{E} respectively). The purple line is the sum of the shown cosine waves, depicting the nature of the Fourier transform.}
  \label{fig:graphical2}
\end{figure*}

\section{Methods}

\subsection{Data}
We trained on the mdCATH dataset of 5,398 monomer domain MD trajectories simulated at 5 temperatures (320K, 348K, 379K, 413K, 450K) with 5 repeats each. Each trajectory consists of up to 450 structures over 1 ns timesteps. We use the standard 80~/~10~/~10 train~/~val~/test split set out in related work~\cite{jingAlphaFoldMeetsFlow2024c,kapusniakMarSFMGenerativeModeling2026}. The test set was curated with mmseqs2~\cite{steineggerMMseqs2EnablesSensitive2017} so that no sequence holds $>20\%$ sequence similarity within the test set, yielding splits with 4304~/~538~/~495 domains respectively. We denote the dataset $\mathcal{D}$ of $N$ training trajectories
\begin{align}
    \mathcal{D} = \left\{\mathbf{X}^{(n)} \in \mathbb{R}^{T_n \times L_n \times 3}  \right\}_{n=1}^N,
\end{align}
where $\mathbf{X}^{(n)}_{t,i} \in \mathbb{R}^3$ denotes the Cartesian coordinates of the $i$-th residue's $C_\alpha$ atom at time $t$. For each $\mathbf{X}^{(n)}$ we sample random, contiguous temporal windows and crop random, contiguous residue subsets
\begin{align}
    \mathbf{X}_{s:s+\tau-1,I} \in \mathbb{R}^{\tau \times \ell \times 3}, \quad &s \sim U\{0,...,T_n-\tau\}, \notag \\ &I \subset \{1,...,L_n\}, \; |I| = \ell,
\end{align}
with $\tau = 256$ and $\ell = 576$. Sequence-based cropping is done after sampled trajectory windows are aligned using a two-stage iterative rigid-core alignment strategy to the native structure detailed in Algorithm~\ref{alg:alignment} to reduce global rotations and translations whilst still retaining large relative subunit motion (Appendix~\ref{sec:appendix_alignment}).

\subsection{Spectral Transformations}
We denote the input reference structure for any given trajectory $\mathbf{X}_\text{ref}^{(n)} \in \mathbb{R}^{L_n \times 3}$ and compute the displacement trajectory for each frame in the sampled window:
\begin{equation}
    \Delta \mathbf{X}_{s+t,i,c}^{(n)} = \mathbf{X}_{s+t,i,c}^{(n)} - \mathbf{X}_{\text{ref},i,c}^{(n)}
\end{equation}

Given a displacement trajectory $\Delta \mathbf{X}_{s:s+\tau-1,I}$, we apply the orthonormal DCT-II transformation along the time dimension. For each residue $i \in I$ and coordinate $c \in \{x,y,z\}$, the transform is given by
\begin{align}
    \mathbf{Z}_{k,i,c}^{(n)}
    =& \alpha_k \sum_{t=0}^{\tau -1}
      \Delta \mathbf{X}_{s+t,i,c}^{(n)}
      \cos \left[ \frac{\pi}{\tau}\Big(t + \frac{1}{2}\Big)k\right], \\
    \alpha_k =&
    \begin{cases}
        \sqrt{1/\tau}, & k=0,\\[2pt]
        \sqrt{2/\tau}, & k>0,
    \end{cases}
\end{align}
where $\alpha_k$ make the transform orthonormal, yielding $k \in \{0,...,\tau-1\}$ strictly real-valued frequency coefficients 
\begin{equation}
    \mathbf{Z}^{(n)} \in \mathbb{R}^{\tau \times \ell \times 3}.
\end{equation}

The DCT avoids DFT-induced boundary discontinuities by implicitly extending the signal with mirrored symmetry. Under this extension, the endpoint $\mathbf \Delta X^{(n)}_{s+\tau -1,i,c}$ connects continuously to its reflection, eliminating the wrap-around discontinuity, making it better suited to non-equilibrium dynamics such as high-temperature unfolding~\cite{yaroslavskyDFTDCTMDCT}.

It is standard in the field of electrical engineering to refer to the first (lowest) frequency of the spectral volume $k=0$ as the Direct Current (DC) component. The DC component represents the per-residue per-channel mean over the trajectory, hence spectral transformation of displacements rather than absolute coordinates means the DC component encodes the mean displacement over the trajectory as opposed to the mean structure. 

Going forward, truncation of the spectral volume will refer specifically to setting all frequency coefficients $>k$ to 0.

\subsection{Diffusion on Spectral Volumes}
\label{sec:spec_diffusion}

\begin{figure}[t]
  \centering
  \includegraphics[width=0.95\columnwidth]{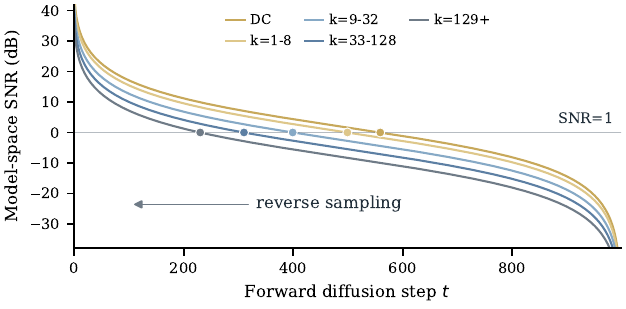}
  \caption{The log-SNR shifted cosine noise schedule enforces hierarchical low to high frequency denoising. Model-space SNR (dB) is the SNR after applying frequency normalisation.}
  \label{fig:cosine_schedule}
\end{figure}

\paragraph{Schedule and noise.}
For a clean spectral target $\mathbf{Z}_0$ the forward diffusion process is defined 
\begin{equation}
    \mathbf{Z}_t =
    \sqrt{\bar{\alpha}_t}\,\mathbf{Z}_0
    + \sqrt{1-\bar{\alpha}_t}\,(\mathbf{w}\odot\boldsymbol{\epsilon}),
    \qquad
    \boldsymbol{\epsilon}\sim\mathcal{N}(\mathbf{0},\mathbf{I}).
\end{equation}
Here \(\bar{\alpha}_t\) is given by a log-SNR-shifted cosine schedule~\cite{nicholImprovedDenoisingDiffusion2021}. The vector \(\mathbf{w}\) contains unit-RMS anisotropic noise multipliers computed from the train-set frequency-scale vectors used for spectral volume normalisation (Appendix~\ref{sec:appendix_normalisation}), with anisotropy strength \(\gamma=0.5\) (Appendix~\ref{sec:appendix_training}). This preserves the total noise power of isotropic diffusion while separating the effective log-SNR trajectories across frequency groups (Figure~\ref{fig:cosine_schedule}). In effect it enforces a hierarchical low to high frequency denoising, low frequencies denoise first setting the global shape and dynamics whilst high frequencies encode atomistic detail akin to how recent diffusion approaches proceed~\cite{chuAllatomProteinGenerative2024,jingEigenFoldGenerativeProtein2023a}.

\paragraph{Objective.}
All models use $\mathbf{Z}_0$-prediction. As the frequency coefficients are the prediction target the model directly reasons on dynamics. Our training loss is a masked $\mathbf{Z}_0$-MSE plus a curriculum scheduled auxiliary detailed in Appendix~\ref{sec:appendix_training}:
\begin{align}
  \mathcal{L}_{\mathrm{MSE}} &=
  \frac{\sum_l m_l\,\|\hat{\mathbf Z}_{0,l} - \mathbf Z_{0,l}\|_2^2}{\sum_l m_l}, \\
  \mathcal{L} &= \mathcal{L}_{\mathrm{MSE}} + \lambda_{\mathrm{aux}}\,\mathcal{L}_{\mathrm{aux}} .
\end{align}

\paragraph{Model.}
Our model, DynaMode, is a spectral convolution architecture that combines a frequency-band mixing full-spectrum trunk with a dedicated low-frequency amplitude-calibration branch. The trunk is inspired by Fourier Neural Operators (FNOs)~\cite{liFourierNeuralOperator2021} and the low-frequency specialist is designed to boost the accuracy of the dominant DC / low-\(k\) amplitudes that encode shape and flexibility. DynaMode achieves rapid sampling time compared to full self-attention transformers by splitting a full linear frequency mixing operation across the spectral volume into blocks that respect approximate timescale groupings (Table~\ref{tab:freq_groups}). Appendix~\ref{sec:appendix_architectures} gives the full architectural specification.

\begin{figure*}[!t]
  \centering
  \includegraphics[width=\textwidth]{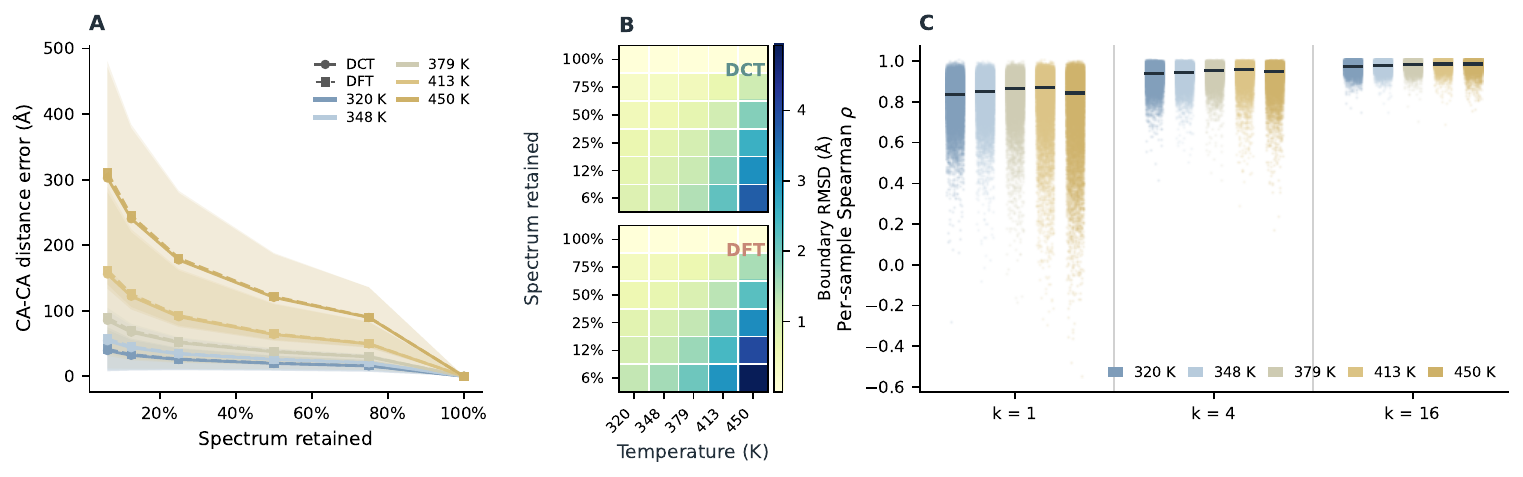}
  \caption{Spectral transformation with DCT followed by truncation leads to significant reconstruction errors. \textbf{A} $C_\alpha-C_\alpha$ distances break down with minimal spectral truncation and this scales with simulation temperature. \textbf{B} While spectral truncation destroys structural validity, DCT (top) is more robust to this than DFT (bottom) at the trajectory boundaries (first and last 5 frames). \textbf{C} A strong positive relationship between RMSF as a measure of flexibility and DCT spectral amplitude of the lowest $k$ frequencies exists when reported by per-sample Spearman correlation ($\rho$).}
  \label{fig:spectral_reconstruction_rmsf}
\end{figure*}

\section{Experiments}

\subsection{Spectral Truncation Destroys Structural Validity}
\label{sec:spectral_error}
Spectral transformations are commonly used in video compression where the highest $>k$ frequencies are discarded or quantized, generally encumbering negligible losses in video fidelity~\cite{wallaceJPEGStillPicture1992}. Considering how MD trajectories could be similarly compressed, we first explored how truncating the spectral volume down to the lowest $k$ modes can effect reconstruction quality through a number of structural validity and dynamics metrics. Specifically, for the set of $k=b/K, \quad b \in \{0.0625, 0.125, 0.25, 0.5, 0.75, 1.0\}$ lowest frequency fractions of the full volume\footnote{We use fractions of the total number of frequencies as the DFT transform has half the number of coefficients as DCT.}, we perform both the DFT and DCT transforms on the trajectory for comparison, zero the frequencies $>k$, before inverse transforming to give the reconstructed trajectories. We aggregated the following metrics over the training set: 
\begin{enumerate}[leftmargin=*, itemsep=0.1pt, topsep=0.1pt]
  \item \textbf{RMSF Spearman}: Measures per-residue flexibility consistency as a proxy for consistency of dynamics.
  \item \textbf{Backbone RMSD}: Measures deviation from the original trajectory giving a direct error measurement.
  \item \textbf{Neighbouring C$_\alpha$-C$_\alpha$ Distances}: Physical validity.
\end{enumerate}

We compared DFT and DCT across temperatures and specifically considered trajectory boundaries (first and last 5 frames) at each temperature to explore how DCT improves upon DFT in the non-equilibrium setting (Figure~\ref{fig:spectral_reconstruction_rmsf}). Interestingly, truncation of the spectral volume had negligible effect on RMSF, (mean 0.96 RMSF spearman at $k=16$ for DCT and $k=8$ for DFT) evidencing the collection of dynamics in the lowest frequencies (Appendix Table~\ref{tab:dct_dft_truncation_errors}). 

We expected DCT to be more robust to error, especially at the boundaries as it does not suffer the same Gibbs ringing phenomenem of DFT on non-equilibrium dynamics after truncation of the high frequencies~\cite{gottliebGibbsPhenomenonIts1997}. Indeed, we found that DCT was marginally more accurate than DFT at trajectory boundaries (Figure~\ref{fig:spectral_reconstruction_rmsf}B), motivating its use throughout this work.

\subsection{Spectral Amplitude Captures Residue Flexibility}
\label{sec:rmsf_spectral_power}
Given the assumption that the low frequencies represent slow collective motion, we next asked how well these low frequency modes represent per-residue flexibility by comparing directly to RMSF. We compute the spectral power for each residue from the lowest $k$ Fourier frequencies $f$ as the squared $\ell_2$-norm of those frequencies
\begin{equation}
    \|f_Z(k)\|_2^2 = \sum_{m=1}^{k} \sum_{c \in \{x,y,z\}} |Z_{c,m}|^2.
\end{equation}
With displacements from native $d_t = x_t - x_{\mathrm{nat}}$, Parseval's identity for the orthonormal DCT states that the per-residue trajectory variance about its mean position $\mathrm{RMSF}_i^2(\mathbf{X})$, equals the non-DC frequencies' summed spectral power,
\begin{equation}
    \mathrm{RMSF}_i^2(\mathbf{X}) \;=\; \tfrac{1}{\tau}\sum_{m=1}^{\tau-1}\sum_{c}|Z_{i,c,m}|^2 \;=\; \tfrac{1}{\tau}\,\|f_{Z_i}(\tau{-}1)\|_2^2,
\end{equation}
so $\tfrac{1}{\tau}\|f_{Z_i}(k)\|_2^2$ is a truncated $\mathrm{RMSF}^2$ that converges to the full quantity as $k \to \tau{-}1$. Because slow modes carry the bulk of the variance, This is already a good proxy given low $k$. The DC term is orthogonal to flexibility as $\sum_c|Z_{i,c,0}|^2/\tau = \|\bar d_i\|^2$ is the squared mean displacement from native. Thus, the per-residue mean-square distance from native decomposes as $\|\bar d_i\|^2 + \mathrm{RMSF}_i^2$. 

As the non-DC frequencies encode the temporal scale of motion (slow hinge vs.\ fast inter-frame jitter) and its direction in $\mathbb{R}^3$, the spectral amplitude is more expressive than RMSF. This decomposition is the primary inductive bias of the spectral target - a model trained to reproduce low-$k$ spectral volumes is by construction reasoning over RMSF.

Analysis of the training set shows a strong positive correlation (per-sample spearman $p$) between per-residue RMSF and DCT amplitude profiles even down to the lowest $k=4$ non-DC frequencies (Figure~\ref{fig:spectral_reconstruction_rmsf}B). Thus, accurate prediction of the lowest non-DC frequencies is the most important prediction task for learning protein flexibility in this regime. This is our primary motivation for the use of a dedicated low-frequency residual correction head in the model (Section~\ref{sec:appendix_architectures}) and $x_0$ prediction over noise or $v$ prediction.

\begin{figure}[t]
  \centering
  \includegraphics[width=0.95\columnwidth]{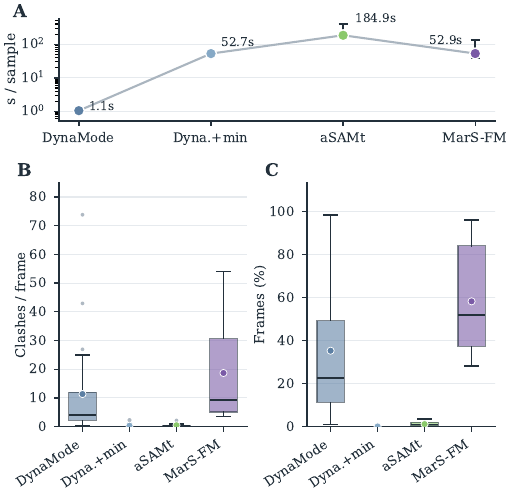}
  \caption{\textbf{A} Median inference time per generated 500-frame trajectory on the same GH200 hardware, with error bars showing the interquartile range. Bottom: Structural-validity distributions across targets, reporting nonbonded $C_\alpha$--$C_\alpha$ clashes below $3.5\,\text{\AA}$ per frame (\textbf{B}) and the percentage of frames containing a nonlocal backbone trace distance below $1.0\,\text{\AA}$ (\textbf{C}).}
  \label{fig:direct_comp}
\end{figure}

\begin{table*}[!t]
  \caption{Trajectory benchmark on mdCATH (all five temperatures, 320--450\,K). MDGen, AlphaFlow-MD and Tempo competitor numbers from published evaluations. $\uparrow$: higher is better. $\downarrow$: lower is better. \textbf{Bold}: best per column among non-oracle methods. Oracle uses the held-out MD trajectory as prediction. Values are medians over 2475 test trajectories.}
  \label{tab:mdcath_trajectory_v12a_v2}
  \begin{center}
    \begin{small}
      \begin{sc}
      \setlength{\tabcolsep}{4pt}
        \begin{tabular}{lccccccc}
        \toprule
        Method & Pair.\ RMSD $r$ $\uparrow$ & Global RMSF $r$ $\uparrow$ & RMWD $\downarrow$ & PCA $\mathcal{W}_2$ $\downarrow$ & PC-sim $\uparrow$ & Weak J $\uparrow$ & Trans.\ J $\uparrow$ \\
        \midrule
        Oracle & 0.992 & 0.885 & 3.08 & 2.21 & -- & 0.822 & 0.482 \\
        \midrule
        MDGen & 0.710 & 0.670 & \textbf{3.36} & 2.62 & 17.19\% & 0.410 & 0.200 \\
        AlphaFlow-MD & 0.410 & 0.410 & 5.62 & 2.38 & \textbf{21.88\%} & 0.420 & \textbf{0.270} \\
        Tempo & 0.770 & 0.670 & 4.21 & \textbf{2.33} & 7.81\% & 0.430 & 0.200 \\
        \midrule
        DynaMode & \textbf{0.854} & \textbf{0.844} & 4.12 & 2.78 & 17.13\% & \textbf{0.620} & 0.246 \\
        \bottomrule
        \end{tabular}
      \end{sc}
    \end{small}
  \end{center}
  \vskip -0.1in
\end{table*}

\begin{table*}[!t]
  \caption{Trajectory benchmark on ATLAS (300\,K). MDGen, AlphaFlow-MD and Tempo competitor numbers from published evaluations. Note that MDGen, Alphaflow and TEMPO each train on a subset of the ATLAS dataset whilst ours is a more difficult out-of-distribution test using the same test set. $\uparrow$: higher is better. $\downarrow$: lower is better. \textbf{Bold}: best per column among non-oracle methods. Oracle uses the held-out MD trajectory as prediction. Values are medians over test targets.}
  \label{tab:atlas_trajectory_v12a_v2}
  \begin{center}
    \begin{small}
      \begin{sc}
      \setlength{\tabcolsep}{4pt}
        \begin{tabular}{lccccccc}
        \toprule
        Method & Pair.\ RMSD $r$ $\uparrow$ & Global RMSF $r$ $\uparrow$ & RMWD $\downarrow$ & PCA $\mathcal{W}_2$ $\downarrow$ & PC-sim $\uparrow$ & Weak J $\uparrow$ & Trans.\ J $\uparrow$ \\
        \midrule
        Oracle & 0.835 & 0.910 & 1.85 & 1.25 & -- & 0.720 & 0.52 \\
        \midrule
        MDGen & 0.480 & 0.500 & 2.69 & 1.89 & 10\% & 0.510 & \textbf{0.410} \\
        AlphaFlow-MD & 0.480 & 0.600 & 2.61 & 1.52 & 44\% & 0.620 & 0.290 \\
        Tempo & \textbf{0.910} & \textbf{0.890} & \textbf{1.49} & \textbf{0.60} & \textbf{76\%} & \textbf{0.740} & 0.380 \\
        \midrule
        DynaMode & 0.665 & 0.734 & 2.65 & 1.72 & 4\% & 0.491 & 0.225 \\
        \bottomrule
        \end{tabular}
      \end{sc}
    \end{small}
  \end{center}
  \vskip -0.1in
\end{table*}

\subsection{Spectral Diffusion Effectively Learns Protein Dynamics}
\label{sec:evaluation}
We trained our diffusion model, DynaMode, to generate full DCT spectral volumes for $\tau=256$ frame windows, given an input monomer structure and temperature, which are inverse transformed into 256 temporally ordered structures.

\paragraph{Evaluation protocol.}
We evaluated DynaMode with a comprehensive set of trajectory evaluation benchmarks  used in recent generative protein dynamics work to assess both ensemble properties and structural validity. Specifically, we follow the method described by~\cite{jingAlphaFoldMeetsFlow2024c} and used in AlphaFlow~\cite{jingAlphaFoldMeetsFlow2024c}, MDGen~\cite{jansonDeepGenerativeModeling2025}, and TEMPO~\cite{xuTEMPOTemporalMultiscale2025b} with the same test splits for comparison with their reported results:
\begin{enumerate}[leftmargin=*, itemsep=-5pt, topsep=-5pt]
    \item The held-out mdCATH test set (495 domains across 5 temperatures, 320--450\,K).
    \item An 82 domain subset of the ATLAS dataset~\citep{vandermeerscheATLASProteinFlexibility2024b} of equilibrium MD simulations at 300\,K (100\,ns trajectories sliced to 1\,ns resolution), serving as an out-of-distribution test at a temperature below the training range.
\end{enumerate}

For a detailed description of the protocol and metrics used we refer the reader to Appendix~\ref{sec:appendix_evaluation}. 

\paragraph{DynaMode is Fast but Suffers from Steric Clashes}
Given reported inference times are not fairly comparable across systems, and other models do not report structural validity metrics, we ran a small direct comparison against the recent ensemble samplers aSAMt and MarS-FM comparing inference times and structural validity before the full test set benchmark. As aSAMt uses a different train/test split, we used the 6 domain (24 trajectories) union of our mdCATH test sets. For each model we adhered to their respective inference protocols and evaluated structural validity according to the distributions of nonbonded $C_\alpha - C_\alpha$ distances in the generated trajectories (Appendix~\ref{sec:appendix_structural_validity}). 

DynaMode is 2 orders of magnitude faster than the next best performing model aSAMt (Figure~\ref{fig:direct_comp}A) on the same GH200 GPU. However, this result highlighted poor structural validity across generated structures in our base approach which, like aSAMt, motivated the use of a brief post-inference energy minimsation~\cite{jansonDeepGenerativeModeling2025} (detailed in Section~\ref{sec:appendix_energy_min}) which we found resolved most steric clashes (Figure~\ref{fig:direct_comp}B\&C) but dampened our superior inference times (Figure~\ref{fig:direct_comp}A). Although our comparison shows validity superior to MarS-FM and aSAMt after energy minimisation, the size of the dataset used limits the power of the comparison. 

\begin{figure*}[!t]
  \centering
  \includegraphics[width=\textwidth]{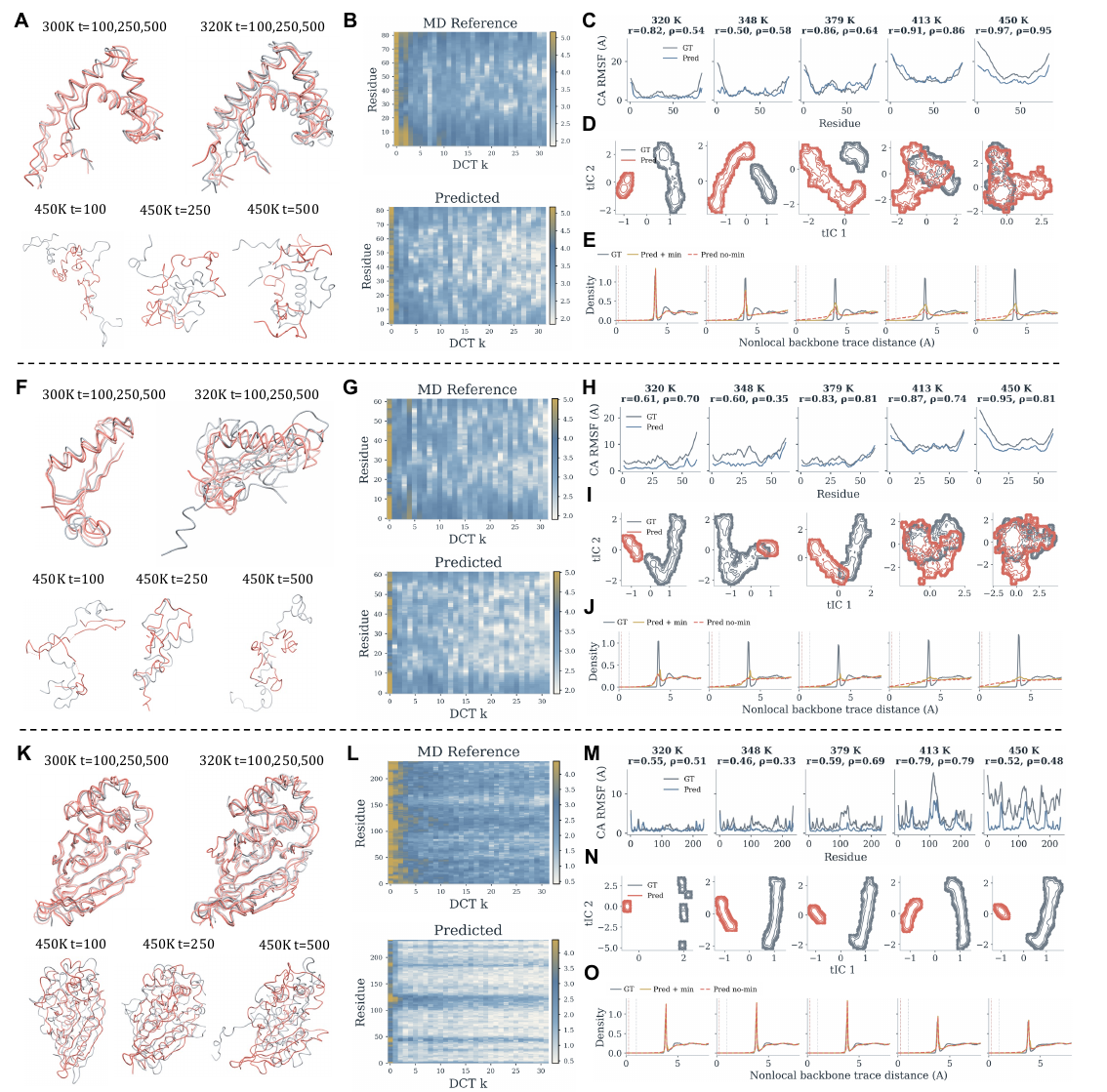}
  \caption{We show three case study domains for DynaMode: 1. (top) mdCATH domain ID 1aabA00 is an 83 residue chain from the train set. 2. (middle) mdCATH domain ID 2fm7A00 is a small 62 residue unfolder from the validation set. 3. (bottom) mdCATH domain ID 4c23B01 from the test set is a 234 residue globular protein. Predicted trajectories were post-inference energy minimised. For each we show structures representing frames 100, 250 and 500 from the predicted (red) and reference MD trajectories (grey) at 300K and 320K overlayed and aligned on each other with decreasing transparency with frame index, and for 450K separately (\textbf{A,F,K}). For 300K we show only the native input PDB structure in grey as a reference MD trajectory is not available at 300K for these mdCATH domains. \textbf{B,G,L} Heatmaps of the first $k=32$ lowest per-residue frequency amplitudes ($\ell_2$-norm over x,y,z channels) of the MD reference and predicted spectral volumes. For each of the mdCATH set temperatures 320K, 348K, 379K, 413K and 450K RMSF plots with Pearson $r$ and Spearman $\rho$ show the predicted trajectories against the MD reference (\textbf{C,H,M}). Similarly tICA free energy overlays at each temperature (\textbf{D,I,N}) and non-local backbone trace distance distributions for the MD reference, predicted with and without energy minimisation (\textbf{E,J,O}) are shown. tICA folding free energy, RMSF and steric clashes calculations are defined in appendix~\ref{sec:appendix_evaluation}.}
  \label{fig:results}
\end{figure*}

\paragraph{Benchmark Results on the mdCATH held-out test set}
We next report performance across the entire mdCATH test set of 495 domains (Table~\ref{tab:mdcath_trajectory_v12a_v2}). Due to computational limits and TEMPO not providing code, the reported results of competitor models are their own published values. We also do not perform post-inference energy minimsation but note that the relaxation step has negligible effect on RMSF. aSAMt is excluded as it used a different train/test split to the other models.

DynaMode achieves the strongest performance on the mdCATH test set in pairwise RMSD $r$ and global RMSF $r$, evidencing the effectiveness of our DCT representation in capturing dynamics (Table~\ref{tab:mdcath_trajectory_v12a_v2}) even in non-equilibrium settings like unfolding (Figure~\ref{fig:results}A). Although MarS-FM~\cite{kapusniakMarSFMGenerativeModeling2026} uses the same splits and benchmark, they report 320K and 450K stratified results on mdCATH only, which we match in Section~\ref{sec:appendix_temp_mdcath_eval}, showing that they narrowly outperform us on all metrics at both temperatures.

\paragraph{}
We present three case study domains, 1aabA00, 2fm7A00 and 4c23B01 where 4c23B01 is an example of a particularly poor prediction, clearly visible from the difference in spectral volumes (Figure~\ref{fig:results}L). The significantly lower predicted amplitudes correspond to reduced flexibility and conformational space misalignment (Figure~\ref{fig:results}N), which aligns with the remarkable consistency in nonlocal backbone trace distance distributions with the the reference MD, particularly compared to 1aabA00 (Figure~\ref{fig:results}E \& J). This highlights the trade off between dynamics and validity, where conservative spectral coefficient predictions (near zero DCT amplitudes) represent model collapse into the input structure, and confident predictions with high amplitudes reproduce dynamics but unless accurately composed fail to maintain validity.

\paragraph{Benchmark Results on the ATLAS dataset}
Table~\ref{tab:atlas_trajectory_v12a_v2} reports trajectory benchmark results on the ATLAS test set. We emphasize that, whilst the performance is significantly worse on the ATLAS test set, this is an out-of-distribution test set at 300K. MDGen, Alphaflow-MD and TEMPO each train on a subset of the ATLAS dataset.

\section{Discussion}
DynaMode is capable of producing temporally coherent $C_\alpha$-only trajectories for single chain monomers under 576 residues at temperatures from 300-450K with a given structure as input. 256 temporally ordered structures are generated in a single diffusion pass in $\sim 1$ seconds per domain on a GH200 GPU beating existing models substantially (Figure~\ref{fig:direct_comp}),
due in part to our spectral convolution architecture (Appendix~\ref{sec:appendix_architectures}) which is significantly more lightweight than a full self-attention transformer with heavy sequence embeddings or pretrained structure modules. Further, diffusion in the spectral domain offers a fundamental inductive bias for learning dynamics due to the relationship between DCT frequencies and RMSF.

By projecting the highly autocorrelated time-domain coordinates onto an orthogonal Fourier basis, we globally decorrelate the temporal signal, eliminating the need for deep autoregressive or expansive convolutions to capture long-range macroscopic state transitions. However, this representation circumvents spatial reasoning which, with imperfect spectral volume predictions, collapses structural geometry (Section~\ref{sec:appendix_structural_validity}). In this regard, ensemble-based methods may outperform~\cite{jingAlphaFoldMeetsFlow2024c,lewisScalableEmulationProtein2025b,jansonDeepGenerativeModeling2025,kapusniakMarSFMGenerativeModeling2026}. We expect this is mostly due to inadequate prediction of the low-mid frequencies (Figure~\ref{fig:results}) which define shape and geometry. This posits the potential for a more elegant architectural combination of spatial and spectral reasoning, which we save for future work. Although we resolve steric clashes with post-inference energy minimisation in a small subset test set, it almost negates our inference speed advantage (Figure~\ref{fig:direct_comp}).

DynaMode is limited to C$_\alpha$ and proteins under 576 residues; other existing methods achieve similar accuracies in all atom regimes~\cite{xuTEMPOTemporalMultiscale2025b,kapusniakMarSFMGenerativeModeling2026,fengBioMDAllatomGenerative2025} at the cost of increased inference times and more complex curriculum training~\cite{xuTEMPOTemporalMultiscale2025b,fengBioMDAllatomGenerative2025}. Although the recent MarS-FM~\cite{kapusniakMarSFMGenerativeModeling2026} achieves higher accuracy compared to DynaMode on ensemble metrics at 320K and 450K on the mdCATH test set (Appendix~\ref{sec:appendix_temp_mdcath_eval}), it does not generate temporally contiguous trajectories and requires construction of trajectory-specific MSMs for training, making it less universal than our DCT approach.

DCT, although more robust to spectral volume truncation than DFT (Figure~\ref{fig:spectral_reconstruction_rmsf}), is still ill-posed for non-equilibrium dynamics due to compensation of non-stationary modes via spectral leakage. Alternative spectral parameterisations such as Chebyshev polynomials~\cite{kondovPolynomialPropagatorsClassical2024,fainDesignOptimalChebyshevexpanded2002} could be explored for such non-periodic directional data. Although we originally hoped for representation compression by truncating the spectral volume to the low frequency modes only, this was shown to destroy physical validity (Appendix~\ref{sec:spectral_error}). However, as is common in video compression, quantisation and/or wavelet transforms of the high frequencies could be explored as a less destructive alternative to truncation~\cite{leeQuantization3DDCTCoefficients1997,bagherizadehStatisticalDCTVector2008}.

Alongside general generative dynamics, DynaMode's low-frequency correction head could be engineered into a self-contained queryable prediction module for per-residue flexibility and dynamics. Indeed, we showed  that the lowest frequency coefficients of DCT transformed displacements from native coordinates are a more expressive measure of residue motion than RMSF which provides only a single scalar average movement value (Section~\ref{sec:rmsf_spectral_power}). We imagine potential applications in rapid structural flexibility prediction of input structures.

It would also be worth exploiting the ameanability of the spectral domain for temporal upsampling by padding the predicted spectral volumes with 0s which, when inverse transformed, increases the resolved frame counts of the trajectories by a smoothing-like interpolation. We emphasise that given the 1ns frame gaps in the training data, the high frequencies do not represent true atomistic and bond vibration timescales as they would using the transform on the full resolution 10ps per-frame ATLAS dataset. Instead, the high frequencies capture inter-frame differences which are artifactual of MD sampling. Either way, the results herein demonstrate the effectiveness of explicit temporal modelling of protein dynamics through spectral transformations.

\section*{Impact Statement}
This work advances machine learning for generative biology by introducing a spectral representation for modelling protein dynamics, together with an architecture designed to accelerate the sampling of molecular dynamics trajectories. By reducing the computational cost of generating dynamical protein ensembles, this work may support downstream applications in biomedical research, protein engineering, and therapeutic development.

The model is trained on publicly available molecular simulation datasets and does not use human-subject data or personally identifiable information. Potential risks include misuse of generative modelling tools for unsupported biological claims, biased conclusions arising from limited simulation coverage, and downstream use in protein design workflows with dual-use implications. These risks are mitigated in part by the coarse-grained C$\alpha$ trajectory setting, the reliance on established scientific software, and the need for substantial additional validation before any experimental or biomedical application.

\section{Code availability}
Model training and inference code, including scripts used for evaluation results, is provided through GitHub (\url{https://github.com/HPuntu/DynaMode}).

\section{Acknowledgements}
The authors acknowledge the use of resources provided by the Isambard-AI National AI Research Resource (AIRR). Isambard-AI is operated by the University of Bristol and is funded by the UK Government’s Department for Science, Innovation and Technology (DSIT) via UK Research and Innovation; and the Science and Technology Facilities Council [ST/AIRR/I-A-I/1023]~\cite{mcintosh-smithIsambardAILeadershipClass2024}. 

The authors would like to acknowledge the use of the University of Oxford Advanced Research Computing (ARC) facility in carrying out this work~\cite{richardsUniversityOxfordAdvanced2015}. 

This work was supported through research funding from the UK Biotechnology and Biological Sciences Research Council (BBSRC) [Grant number EP/S024093/1] and Bayer AG, in collaboration with Bayer AG.

We are grateful to valuable advice from colleagues in the Oxford Protein Informatics Group (OPIG) including Nele Quast for her helpful comments on first drafts.

\section{Competing interests statement}
C.M.D. discloses membership of the Scientific Advisory Board of Fusion Antibodies and AI Proteins
632 as well as being a founder of Dalton. All other authors declare no conflict of interest.

\bibliography{icml_2026}
\bibliographystyle{icml2026}

\newpage
\appendix
\onecolumn
\section{Appendix}

\subsection{Alignment}
\label{sec:appendix_alignment}
All MD trajectories are aligned to the native reference structure using the Kabsch algorithm~\cite{lawrencePurelyAlgebraicJustification2019} which solves for 
\begin{equation}
    \mathbf{R} = \mathbf{U}\,\mathrm{diag}(1,1,\det(\mathbf{U}\mathbf{V}^\top))\,\mathbf{V}^\top,\quad
    \mathbf{U}\mathbf{\Sigma}\mathbf{V}^\top = \mathrm{SVD}(\mathbf{X}^\top \mathbf{Y}).
\end{equation}
We use a two a 2-stage iterative rigid-body alignment where we first perform global alignment, before selecting the lowest 50\% RMSF residues as the rigid core and re-aligning by computing the rotation matrix on the rigid core and applying it to the globally aligned trajectory as detailed in Algorithm~\ref{alg:alignment}.

\begin{algorithm}[htbp]
  \caption{Rigid-Core Trajectory Alignment}
  \label{alg:alignment}
  \begin{algorithmic}
    \STATE {\bfseries Input:} Trajectory $\mathcal{T} = \{\mathbf{X}_t\}_{t=1}^T$ where $\mathbf{X}_t \in \mathbb{R}^{N \times 3}$, Native Structure $\mathbf{X}_{nat}$
    \STATE {\bfseries Output:} Aligned Trajectory $\mathcal{T}'$
    
    \STATE \COMMENT{Step 1: Global Alignment}
    \STATE $\mathbf{X}_{nat} \gets \mathbf{X}_{nat} - \text{mean}(\mathbf{X}_{nat})$ \COMMENT{Mean centering}
    \FOR{$t=1$ {\bfseries to} $T$}
        \STATE $\mathbf{X}_t \gets \mathbf{X}_t - \text{mean}(\mathbf{X}_t)$
        \STATE $\mathbf{R}_t \gets \text{Kabsch}(\mathbf{X}_t^{C_\alpha}, \mathbf{X}_{nat}^{C_\alpha})$ \COMMENT{Compute rotation via SVD}
        \STATE $\mathbf{X}_t \gets \mathbf{X}_t \mathbf{R}_t$
    \ENDFOR
    
    \STATE \COMMENT{Step 2: Rigid Core Identification}
    \STATE Compute RMSF for all $C_\alpha$ atoms: $\rho_i = \sqrt{\frac{1}{T} \sum_{t=1}^T \|\mathbf{x}_{i,t} - \bar{\mathbf{x}}_i\|^2}$
    \STATE Let $\mathcal{S}_{core}$ be the indices of $C_\alpha$ atoms where $\rho_i \leq \text{median}(\rho)$
    
    \STATE \COMMENT{Step 3: Refined Alignment}
    \FOR{$t=1$ {\bfseries to} $T$}
        \STATE $\mathbf{R}'_t \gets \text{Kabsch}(\mathbf{X}_t^{\mathcal{S}_{core}}, \mathbf{X}_{nat}^{\mathcal{S}_{core}})$
        \STATE $\mathbf{X}'_t \gets \mathbf{X}_t \mathbf{R}'_t$
    \ENDFOR
    \STATE \textbf{return} $\mathcal{T}' = \{\mathbf{X}'_t\}_{t=1}^T$
  \end{algorithmic}
\end{algorithm}

\subsection{Conditioned Frequency Normalisation}
\label{sec:appendix_normalisation}
We compute bucketed frequency normalisation factors for each Cartesian channel $c$ of each frequency coefficient $k$ from the full train set of DCT transformed native-displacement $C_\alpha$ spectral volume $z_d$. For each train set trajectory we sample enough windows of size $\tau=256$ to cover the number of frames $T$ typically giving overlapped trajectory windows for each trajectory. For each spectral feature $d=(k,c)$, the reference statistic is the bucket-specific robust amplitude
\begin{equation}
  a_d^{(b)} = Q_{0.75}\!\left(|z_d| \mid b\right),
\end{equation}
where \(b\) is a temperature, size (number of residues), or temperature--size bucket. The deployed scale table uses a shrunk temperature--size estimate
\begin{equation}
  \sigma_d^{(b)} = 0.25\,\sigma_d^{(\mathrm{global})}
  + 0.75\,a_d^{(b)}.
\end{equation}
This shrinkage retains the dominant dependence on temperature and protein size while avoiding a fully independent scale for every bucket and feature.

\begin{figure}[t]
  \centering
  \includegraphics[width=\textwidth]{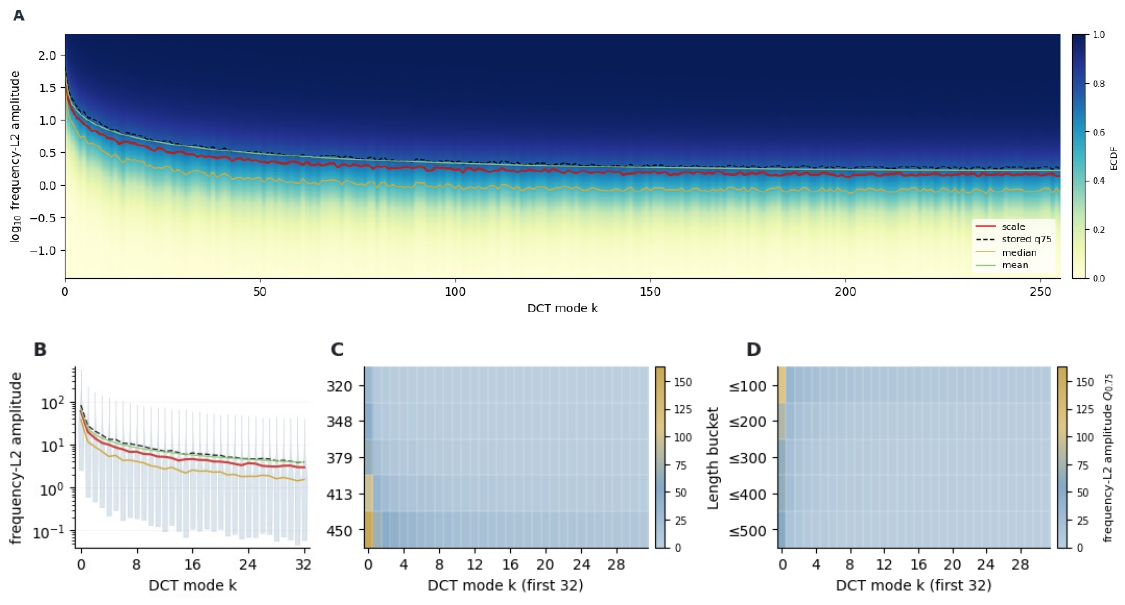}
  \caption{Conditioned DCT frequency-scale statistics used for spectral normalisation. \textbf{A} All-frequency ECDF heatmap over all \(K=256\) DCT modes, with the selected normalisation scale (red line) and stored \(Q_{0.75}\) amplitude (dashed) overlaid. \textbf{B} Low-frequency log-amplitude distribution summary shown as one vertical violin per mode, again with same scale overlaid. \textbf{C} Temperature-conditioned \(Q_{0.75}\) amplitudes for the first 32 modes. \textbf{D} Size-conditioned \(Q_{0.75}\) amplitudes for the first 32 modes.}
  \label{fig:appendix_freq_scales}
\end{figure}

Figure~\ref{fig:appendix_freq_scales} shows the statistical landscape of the train-set DCT spectral volumes. The global DCT amplitude falls steeply from the DC mode and then flattens, while both temperature and size mainly perturb the DC and low-frequency scales. Temperature partitioning has the most pronounced effect on frequency statistics. High-temperature trajectories require much larger low-\(k\) scales, consistent with unfolding drift and slow collective motion. 

To quantify the normalisation effect, we compare the observed bucket statistic \(a_d^{(b)}\) against the scale selected by each scheme using mean absolute log-ratio,
\begin{equation}
  \Delta = \frac{1}{|\mathcal{B}|\,|\mathcal{D}|}
  \sum_{b\in\mathcal{B}}\sum_{d\in\mathcal{D}}
  \left|\log\frac{a_d^{(b)}}{\sigma_d^{(b)}}\right|,
\end{equation}
averaged over the 25 observed temperature/size training buckets. Lower values mean the normalised coefficient amplitudes are closer to a common robust scale across regimes. Table~\ref{tab:appendix_normalisation} shows that temperature conditioning explains most of the improvement over a single global table, while the final shrunk temperature--size table gives the best match, especially for the DC and \(0<k<8\) drift modes.

\begin{table*}[t]
  \caption{Normalisation-scale mismatch across train-set temperature/size buckets. Values are mean absolute log-ratios between observed bucket \(Q_{0.75}(|z|)\) and the scale used by each scheme - lower is better. Bands follow the \(K=256\) DCT frequency grouping used for the normalisation analysis, with the DC mode separated from the remaining drift modes.}
  \label{tab:appendix_normalisation}
  \begin{center}
    \begin{small}
      \setlength{\tabcolsep}{3.5pt}
      \begin{tabular}{lcccccc}
        \toprule
        Scheme & All-\(k\) & DC & \(0<k<8\) & \(8\leq k<32\) & \(32\leq k<128\) & \(128\leq k<256\) \\
        \midrule
        Global & 0.531 & 0.524 & 0.751 & 0.715 & 0.562 & 0.461 \\
        Temperature only & 0.263 & 0.224 & 0.335 & 0.343 & 0.279 & 0.233 \\
        Size only & 0.490 & 0.489 & 0.701 & 0.652 & 0.516 & 0.429 \\
        Temperature\(\times\)size shrunk & 0.137 & 0.147 & 0.220 & 0.201 & 0.145 & 0.114 \\
        \bottomrule
      \end{tabular}
    \end{small}
  \end{center}
  \vskip -0.1in
\end{table*}

Since these scales remove some physically meaningful amplitude variation, the selected low-frequency scales are also passed back to the model through the global scale-conditioning vector described in Section~\ref{sec:appendix_conditioning}. This enables us to preserve some awareness of the magnitude of differences across frequencies which is likely a strong conditioning factor to predict spectral volumes across temperatures and sizes.

\subsection{Residualised DC target.}
We additionally stabilise the most dominant spectral component, the DC term, by subtracting a train-set baseline before diffusion. The DC coefficient corresponds to the mean displacement over the trajectory window, so it carries large systematic offsets associated with temperature, size, and overall structural drift. Asking the model to predict this raw baseline directly wastes capacity on a largely predictable offset. From the train-set we compute a per-protein per-temperature per-residue baseline mean DC component over repeats,
\begin{equation}
  \boldsymbol{\mu}_{\mathrm{DC}}^{(n,\text{temp})} \in \mathbb{R}^{L \times C},
\end{equation}
Before the forward diffusion process we residualise the DC coefficient of the clean spectral volume so that
\begin{equation}
  \mathbf{Z}^{\mathrm{res}}_{l,0,:}
  = \mathbf{Z}_{l,0,:} - \boldsymbol{\mu}_{\mathrm{DC},l,:}^{(n,\text{temp})}.
\end{equation}
The model is therefore trained to predict a residualised low-frequency target rather than the absolute DC coefficient. After denoising, the stored DC baseline is added back before inverse transformation to the time domain. This allows the model to focus on learning residual fluctuation around the expected low-frequency drift, rather than the mean drift itself.

\subsection{Tokenisation}
\label{sec:appendix_tokenisation}
For a window represented by \(K\) retained DCT coefficients and \(C=3\) Cartesian displacement channels, each residue \(l\) is encoded as a flattened spectral token
\begin{equation}
  \tilde{\mathbf{Z}}^{(n)}_l \in \mathbb{R}^{K\times C},
\end{equation}
where \(\tilde{\mathbf{Z}}\) denotes the normalised spectral volume. In practice we use $K=\tau$ but this formulation permits spectral volume truncation by choosing $K<\tau$. A per-frequency linear map is applied independently to each \((k,c)\) slice, projecting into the hidden dimension \(H\). The resulting tensor is flattened to \(D = KH\). We also add a learned frequency embedding before flattening so the model can distinguish DCT modes explicitly. Rotary positional embeddings (RoPE)~\citep{suRoFormerEnhancedTransformer2024} are then applied along the residue axis.

\subsection{Conditioning}
\label{sec:appendix_conditioning}
Native-structure conditioning enters locally at the token input. For each residue and frequency mode we concatenate the normalised spectral coefficient, the native \(C_\alpha\) coordinates (scaled by a constant factor), and residue-level features including amino-acid identity and secondary structure before the per-frequency input projection.

\paragraph{Global Conditioning.}
Global conditioning is injected through AdaLN-Zero~\citep{peeblesScalableDiffusionModels2023}. The conditioning vector
\begin{equation}
  \mathbf{c} = f_{t_d}(t_d) + f_\text{temp}(\text{temp}) + f_s(s) + f_{\mathrm{size}}(L_{\mathrm{eff}})
  + f_{\mathrm{seq}}(\mathbf{a}) + f_{\mathrm{ss}}(\mathbf{q}) + f_{\mathrm{scale}}(\boldsymbol{\sigma})
\end{equation}
combines diffusion time \(t_d\), temperature (\(\text{temp}\)), normalised window position \(s\), effective sequence length \(L_{\mathrm{eff}}\), pooled sequence features \(\mathbf{a}\), pooled secondary-structure features \(\mathbf{q}\), and the selected normalisation-scale features \(\boldsymbol{\sigma}\). We use \(d_{\mathrm{cond}} \ll D\) to decouple conditioning capacity from token width. All AdaLN modulation layers and the final per-frequency output projection are zero-initialised.

\paragraph{Sequence Embedding.}
Amino acid sequence is one-hot encoded and embedded in the model through a single linear layer. We also precompute secondary structure using dssp~\cite{hekkelmanDSSP4FAIR2025} and embed that through another linear layer. Both feature types are used in two ways. First, they are injected locally by concatenating the per-residue embeddings to the spectral token input before the per-frequency projection. Second, they are pooled across valid residues with a masked mean and added to the global conditioning vector \(\mathbf{c}\).

\paragraph{Size conditioning.}
Protein size is discretised into coarse bins matching the those used for the normalisation factors.
\begin{equation}
  L_{\mathrm{eff}} \in \{\leq 100,\leq 200,\leq 300,\leq 400,\leq 500,\leq 600,>600\},
\end{equation}
computed from the masked residue count in the current crop. A learned embedding of this size bin is projected into the global conditioning vector.

\subsection{Architecture}
\label{sec:appendix_architectures}

\paragraph{Spectral Convolution.}
The main body of DynaMode is a spectral-convolution diffusion trunk that predicts the full \(K\)-mode output for every residue. Each block contains (i) masked self-attention across residues and (ii) an FNO-style spectral operator acting along the frequency axis. For (ii), tokens are reshaped to \((B\times L, H, K)\), treating frequency as a one-dimensional domain, and three operations are applied in parallel:
\begin{enumerate}[leftmargin=*, itemsep=0pt, topsep=2pt]
  \item \textbf{SpectralConv1d}: a learned convolution \((H, H, K_{\mathrm{modes}})\) that mixes the lowest \(K_{\mathrm{modes}}\) frequencies.
  \item \textbf{Pointwise Conv1d}: a \(1{\times}1\) pointwise convolution over the coordinate channels.
  \item \textbf{Cross-frequency mixing}: a frequency mixer operating independently within physically motivated frequency groups.
\end{enumerate}
The outputs are summed, passed through a GELU nonlinearity, and reshaped back to \((B, L, D)\). The cross-frequency mixer is \emph{block diagonal} rather than fully dense. For \(K=256\) we partition the frequency bands as follows
\begin{equation}
  [0,8],\; [8,32],\; [32,128],\; [128,256],
\end{equation}
corresponding roughly to drift/slow collective motion, intermediate backbone-scale modes, and high-frequency thermal jitter. This keeps mixing within physically similar spectral regimes and reduces destructive leakage between slow and fast modes.

\paragraph{Low-frequency amplitude calibration head.}
Given the importance of the low frequencies for dynamics representation we boosted the model's capacity to predict them with an amplitude-calibration branch acting as a residual correction on top of the stronger main trunk for a small set of the lowest frequencies. In practice performance was better with a narrower band of low frequencies ($<8$).

Let \(M\) denote the number of target low-frequency modes to recalibrate and \(M_{\mathrm{ctx}} \geq M\) the number of low-\(k\) context modes exposed to the head. The trunk first predicts the full spectral volume \(\hat{\mathbf{Z}}^{\mathrm{trunk}}_{l,k,:}\). We then extract the context tensor
\begin{equation}
  \mathbf{u}_l = \mathrm{vec}\!\left(\hat{\mathbf{Z}}^{\mathrm{trunk}}_{l,\,0:M_{\mathrm{ctx}}-1,:}\right),
\end{equation}
concatenate it with residue-local conditioning features, and process it with a small residue-level transformer to produce per-mode log-gains
\begin{equation}
  \mathbf{g}_{l} = g_{\mathrm{amp}}\!\left(
    \mathbf{u}_{l},
    \mathbf{c},
    \mathbf{e}^{\mathrm{seq}}_{l},
    \mathbf{e}^{\mathrm{ss}}_{l},
    \mathbf{x}^{\mathrm{ref}}_{l}
  \right) \in \mathbb{R}^{M}.
\end{equation}
The gains are applied multiplicatively to the amplitudes of the first \(M\) trunk-predicted vectors while preserving their direction:
\begin{equation}
  \hat{\mathbf{z}}_{l,k} = \hat{\mathbf{Z}}^{\mathrm{trunk}}_{l,k,:}, \qquad
  a_{l,k} = \|\hat{\mathbf{z}}_{l,k}\|_2,
  \qquad
  \mathbf{d}_{l,k} = \hat{\mathbf{z}}_{l,k} / \max(a_{l,k}, \epsilon).
\end{equation}
The calibrated low-frequency prediction becomes
\begin{equation}
  \hat{\mathbf{Z}}^{\mathrm{low}}_{l,k,:}
  =
  \exp(g_{l,k})\, a_{l,k}\, \mathbf{d}_{l,k},
  \qquad k < M,
\end{equation}
while \(\hat{\mathbf{Z}}^{\mathrm{low}}_{l,k,:} = \hat{\mathbf{Z}}^{\mathrm{trunk}}_{l,k,:}\) for \(k \geq M\).

\paragraph{Post-DCT geometry refiner and differentiable SHAKE.}
To account for observed structural collapse and steric clashes we augment the spectral predictor with an explicit coordinate-space geometry module. After denoising, de-normalisation, DC reconstruction, and inverse DCT, the predicted representation is mapped back to absolute \(C_\alpha\) coordinates
\(\hat{\mathbf{X}} \in \mathbb{R}^{T\times L\times 3}\). A lightweight residual refiner \(r_\theta\) then acts directly on these reconstructed coordinates,
\begin{equation}
  \mathbf{X}^{\mathrm{ref}} = \hat{\mathbf{X}} +
  \Delta_\theta(\hat{\mathbf{X}}),
  \qquad
  \|\Delta_{\theta,t,i,:}\|_\infty \leq \Delta_{\max},
\end{equation}
where \(\Delta_\theta\) is a small 1-D convolutional stack over the residue axis applied independently to each frame. The final projection layer is zero-initialised, so the auxiliary learns only a bounded post-iDCT residual correction. We used a refiner with hidden width 32, depth 2, kernel size 5, and \(\Delta_{\max}=0.5\,\text{\AA}\).

Because the DCT inverse is coordinate-wise and does not enforce chain geometry, we follow the refiner with a differentiable SHAKE-style \(C_\alpha\)--\(C_\alpha\) projection~\citep{krautlerFastSHAKEAlgorithm2001}. For every adjacent residue pair at every frame, define
\begin{equation}
  \mathbf{b}_{t,i} = \mathbf{X}^{\mathrm{ref}}_{t,i+1,:} -
  \mathbf{X}^{\mathrm{ref}}_{t,i,:}, \qquad
  r_{t,i} = \max(\|\mathbf{b}_{t,i}\|_2,\epsilon),
\end{equation}
and apply the symmetric update
\begin{equation}
  \boldsymbol{\delta}_{t,i}
  = \frac{1}{2}\left(d^* - r_{t,i}\right)
  \frac{\mathbf{b}_{t,i}}{r_{t,i}}, \qquad
  \mathbf{X}_{t,i,:} \leftarrow \mathbf{X}_{t,i,:} - \boldsymbol{\delta}_{t,i}, \quad
  \mathbf{X}_{t,i+1,:} \leftarrow \mathbf{X}_{t,i+1,:} + \boldsymbol{\delta}_{t,i},
\end{equation}
with target distance \(d^*=3.8\,\text{\AA}\). We iterate this projection for at most 2 passes over valid adjacent pairs. Since the projection remains inside the forward graph, training can expose the spectral trunk and coordinate refiner to the geometry-corrected trajectory used for downstream losses. We also penalise the refiner's SHAKE residual during the geometry-loss phase so that the learned refiner, rather than SHAKE alone, absorbs systematic bond-length corrections.

\subsection{Training}
\label{sec:appendix_training}

We train on randomly sampled 256-frame windows translating to a 256-mode DCT. Sequences are cropped to 576 residues. Distributed training across 4 GH200 GPUs allows for a batch size of 200. We trained for 300 epochs with AdamW, using a peak learning rate of \(3\times 10^{-5}\) taking 24 hours in total. Weight decay is applied in the standard decoupled form with coefficient \(0.05\) on matrix weights and no decay on bias or normalisation parameters. Optimisation used a OneCycle learning-rate schedule with cosine annealing. As the schedule sets \(\texttt{pct\_start}=\min(0.1,5/\texttt{epochs})\), this corresponds to a five-epoch warmup in the final 300-epoch run. We clip the global gradient norm to \(1.0\), and use bfloat16 mixed precision.

\paragraph{Noise Schedule}
We sample from 1000 diffusion timesteps each training step using a log-SNR-shifted cosine schedule for the scalar \(\bar{\alpha}_t\). Spectral anisotropy is applied in the flattened (frequency,coordinate) feature dimension \(d=(k,c)\), using the same train-set frequency-scale vector \(\sigma_d^{(b)}\) used for spectral normalisation in bucket \(b\). The unnormalised and unit-RMS noise multipliers are
\begin{equation}
    \tilde{w}_d =
    \left(
        \frac{\sigma_d^{(b)}}{\min_j \sigma_j^{(b)}}
    \right)^\gamma,
    \qquad
    w_d =
    \frac{\tilde{w}_d}
    {\sqrt{|\mathcal{F}|^{-1}\sum_{j\in\mathcal{F}}\tilde{w}_j^2}},
    \qquad
    \gamma=0.5,
\end{equation}
where \(\mathcal{F}\) is the set of flattened spectral features. Since the largest scales occur in the DC and low-\(k\) modes, this gives low-frequency trajectory components proportionally stronger corruption in both forward diffusion and the initial sampler noise while preserving unit-RMS total noise power.

Algorithm~\ref{alg:training} summarizes the training loop. The implementation supports \(\mathbf{Z}_0\)-, noise-, and \(v\)-prediction targets, but all reported DynaMode models use \(\mathbf{Z}_0\)-prediction. The clean target is the normalised, optionally DC-residualised DCT spectral volume of the native-frame displacement trajectory. We employ Classifier Free Guidance (CFG)~\cite{hoClassifierFreeDiffusionGuidance2022a} dropout on temperature and structural conditioning with 15\% probability.

\begin{algorithm}[htbp]
  \caption{Spectral Diffusion Training}
  \label{alg:training}
  \begin{algorithmic}
    \STATE {\bfseries Input:} Training set \(\mathcal{D}\), model \(f_\theta\), transform pipeline \(\Phi\), diffusion process \(q\), epochs \(E\), retained modes \(K\), window length \(T=256\), optimizer \(O\), scheduler \(H\)
    \STATE {\bfseries Output:} Best checkpoint \(\theta^\star\)

    \STATE Initialise AdamW with decoupled decay on matrix weights only, OneCycle scheduler, and best validation score \(b\gets-\infty\)
    \FOR{epoch \(e=1\) {\bfseries to} \(E\)}
        \STATE Enable random temporal-window sampling in the training dataset
        \FOR{minibatch \(\mathcal{B}\) from the training loader}
            \STATE Load aligned trajectory windows \(\mathbf{X}\), native structures \(\mathbf{X}_{\mathrm{nat}}\), temperatures \(T_\text{temp}\), residue masks \(\mathbf{m}\), residue features \(\mathbf{a}\), DSSP features \(\mathbf{q}\), and optional torsion features
            \STATE Form coordinate representation \(\mathbf{Y}=R(\mathbf{X},\mathbf{X}_{\mathrm{nat}})\) and concatenate torsion channels when present
            \STATE Jitter temperatures during training and compute \(\bar{T}_\text{temp}=\mathrm{clip}((T_\text{temp}-250)/200,0,1)\)
            \STATE Sample classifier-free conditioning dropout mask \(\mathbf{d}\)
            \STATE \(\tilde{\mathbf{Z}}_0 \gets \Phi(\mathbf{Y}\odot\mathbf{m}; K)\) (DCT, truncation, frequency normalisation)
            \STATE Residualise the DC coefficient of \(\tilde{\mathbf{Z}}_0\) using per-residue or bucket-level baselines when available
            \STATE Sample diffusion steps \(t\) and noise \(\boldsymbol{\epsilon}\); compute \(\tilde{\mathbf{Z}}_t=q(\tilde{\mathbf{Z}}_0,t,\boldsymbol{\epsilon})\)
            \STATE Build residue/channel mask \(\mathbf{M}\) and, for hierarchical runs, sample visible and target frequency groups
            \STATE \(\hat{\mathbf{u}}\gets f_\theta(\tilde{\mathbf{Z}}_t,t,\bar{T_\text{temp}},\mathbf{X}_{\mathrm{nat}},\mathbf{a},\mathbf{q},\mathbf{m},\mathbf{d})\)
            \STATE Choose target \(\mathbf{u}\): \(\tilde{\mathbf{Z}}_0\) for \(\mathbf{Z}_0\)-prediction, \(\boldsymbol{\epsilon}\) for noise-prediction, or \(v_t\) for \(v\)-prediction
            \STATE \(\displaystyle
              \mathcal{L}_{\mathrm{denoise}} \gets
              \frac{\sum \mathbf{M}\,w_k\,w_t\,\|\hat{\mathbf{u}}-\mathbf{u}\|_2^2}
                   {\sum \mathbf{M}\,w_k\,w_t + \epsilon}\)
            \STATE Recover \(\hat{\tilde{\mathbf{Z}}}_0\) from \(\hat{\mathbf{u}}\) when auxiliary clean-target losses are active
            \STATE Add scheduled auxiliary losses: spectral amplitude, signed low-\(k\), DC, and IDCT-space geometry/refiner/SHAKE losses
            \STATE \(\mathcal{L}\gets\mathcal{L}_{\mathrm{denoise}}+\mathcal{L}_{\mathrm{aux}}\)
            \IF{\(\mathcal{L}\) or gradients are non-finite}
                \STATE Skip the optimizer update on all distributed ranks
            \ELSE
                \STATE Backpropagate \(\mathcal{L}\), clip global gradient norm to \(1.0\), step \(O\), and step \(H\)
            \ENDIF
        \ENDFOR
        \STATE Disable random validation windows and evaluate one-step loss plus full inference metrics
        \STATE Save the latest checkpoint
        \STATE \(s\gets \mathrm{RMSF\ Spearman}+\mathrm{LDDT}\)
        \IF{\(s>b\)}
            \STATE \(b\gets s\); save \(\theta^\star\) as the best inference checkpoint
        \ENDIF
    \ENDFOR
    \STATE \textbf{return} \(\theta^\star\)
  \end{algorithmic}
\end{algorithm}

The training objective combines a masked \(\mathbf{Z}_0\)-MSE loss with auxiliary low frequency spectral amplitude and geometry terms:
\begin{align}
    \mathcal{L}_{\mathrm{MSE}} &=
    \frac{\sum_l m_l\,\|\hat{\tilde{\mathbf{Z}}}_{0,l}-\tilde{\mathbf{Z}}_{0,l}\|_2^2}
         {\sum_l m_l+\epsilon}, \\
    \mathcal{L} &= \mathcal{L}_{\mathrm{MSE}}
    + \lambda_{\mathrm{amp}}\,\mathcal{L}_{\mathrm{amp}}
    + \lambda_{\mathrm{geo}}\,\mathcal{L}_{\mathrm{geo}}.
\end{align}
\begin{align}
  \mathcal{L}_{\mathrm{amp}} &=
  \frac{\sum_{l,k} m_l\,\mathrm{Huber}_\delta\!\bigl(A_{l,k}(\hat{\tilde{\mathbf{Z}}}_0) - A_{l,k}(\tilde{\mathbf{Z}}_0)\bigr)}{K \sum_l m_l}.
\end{align}

The spectral amplitude loss \(\mathcal{L}_{\mathrm{amp}}\) matches per-residue, per-frequency magnitudes
\begin{equation}
    A_{l,k}(\tilde{\mathbf{Z}}) = \big\|\tilde{\mathbf{Z}}_{l,\,3k:3k+3}\big\|_2,
\end{equation}
between the predicted clean representation \(\hat{\tilde{\mathbf{Z}}}_0\) and ground truth \(\tilde{\mathbf{Z}}_0\) using a Huber loss (\(\delta=0.5\)). This preserves the full \(K\)-dimensional amplitude profile per residue, encouraging the model to capture frequency-dependent motion and helping prevent suppression of low-frequency amplitudes by the $\mathbf{Z}_0$-MSE objective. 

Finally we apply exponential decay frequency weights inside the spectral losses so that the low-frequency modes, which dominate the physical motion and are hardest to calibrate, receive stronger supervision.

\subsection{Inference}
\label{sec:appendix_inference}
The reference structure \(\mathbf{X}^{(n)}_{\mathrm{ref}}\) in the form of $C_\alpha$ coordinates, and temperature, provide the primary input structural conditioning. Generated trajectories are implicitly aligned to this input structure as training trajectories were aligned to their respective input native structures. Classifier-free guidance (CFG) dropout (\(p=0.15\)) is applied to temperature and reference conditioning only. Standard sampling uses $50$ denoising steps; the predicted spectral representation \(\hat{\tilde{\mathbf{Z}}}_0\) is inverse-DCT transformed and decoded to absolute coordinates over time. For structural refinement these coordinates are then passed through the coordinate refiner and differentiable SHAKE projection described above.

Inference is implemented as a windowed trajectory sampler, summarised in Algorithm~\ref{alg:inference}. Given one or more input PDB structures, we extract \(C_\alpha\) coordinates, residue identity features, DSSP features, optional torsion features, and a residue mask. Each requested trajectory is generated as one or more \(T=256\) frame windows. For multi-window trajectories, the sampler can either condition every window on the original native structure or chain windows by conditioning the next window on the final frame of the previous window. The generated windows are concatenated, trimmed to the requested frame count, and optionally passed through a \(C_\alpha\) minimisation pass before export.

Given one or more input PDB structures, we extract \(C_\alpha\) coordinates, residue identity features, DSSP features, optional torsion features, and a residue batch padding mask. \(K=256\) DCT spectral volumes \(\hat{\tilde{\mathbf{Z}}}_0\) are denoised over $50$ steps before inverse transforming into a \(T=256\) frame window which is implicitly aligned to the input structure. For structural refinement these coordinates are then passed through the coordinate refiner and differentiable SHAKE projection described above. Inference can be chained by providing say, the final frame of one prediction, as input, and output trajectories stacked for trajectory sampling beyond $256$ frames. Optional $C_\alpha$ energy minimsation can be performed.
 
\begin{algorithm}[htbp]
  \caption{Windowed Spectral Diffusion Inference}
  \label{alg:inference}
  \begin{algorithmic}
    \STATE {\bfseries Input:} Checkpoint/configuration \(\mathcal{C}\), PDB structures \(\{\mathbf{P}^{(b)}\}_{b=1}^B\), temperatures \(\{T_{\text{temp},b}\}_{b=1}^B\), requested frames \(F\), window length \(T=256\), retained modes \(K\), ODE steps \(N\), CFG scale \(w\)
    \STATE {\bfseries Output:} Generated trajectories \(\{\hat{\mathbf{X}}^{(b)}_{1:F}\}_{b=1}^B\)

    \STATE Build runtime from \(\mathcal{C}\): model \(f_\theta\), diffusion sampler \(S\), spectral transform pipeline \(\Phi\), representation map \(R\)
    \STATE Load checkpoint weights and set \(f_\theta\) to evaluation mode
    \FOR{each input structure \(\mathbf{P}^{(b)}\)}
        \STATE Extract native coordinates \(\mathbf{X}^{(b)}_{\mathrm{nat}}\), topology, residue features \(\mathbf{a}^{(b)}\), DSSP features \(\mathbf{q}^{(b)}\), and mask \(\mathbf{m}^{(b)}\)
        \STATE Optionally compute native torsion features and an NMA RMSF prior from \(\mathbf{X}^{(b)}_{\mathrm{nat}}\)
    \ENDFOR
    \STATE Pad all residue-wise tensors in the batch to \(L_{\max}\)
    \STATE \(J \gets \lceil F/T\rceil\); initialise conditioning structures \(\mathbf{C}^{(b)}_1 \gets \mathbf{X}^{(b)}_{\mathrm{nat}}\)
    \FOR{\(j=1\) {\bfseries to} \(J\)}
        \STATE \(s_j \gets ((j-1)T)/\max(F-1,1)\) \COMMENT{Normalised window-position conditioning}
        \STATE \(D \gets K \cdot C_{\mathrm{model}}\) for DCT spectral models
        \STATE Sample initial latent \(\mathbf{z}_N \sim \mathcal{N}(\mathbf{0},\boldsymbol{\Sigma}_{\mathrm{aniso}})\) with invalid residues and channels masked out
        \STATE Normalise temperatures \(\bar{T}_{\text{temp},b} \gets \mathrm{clip}((\tau_b-250)/200,0,1)\)
        \STATE Wrap \(f_\theta\) with classifier-free guidance scale \(w\)
        \STATE \(\hat{\tilde{\mathbf{Z}}}_{0,j} \gets S_{\mathrm{ODE}}(f_\theta,\mathbf{z}_N,\mathbf{C}_j,\bar{T}_{\text{temp}},s_j,\mathbf{a},\mathbf{q},\mathbf{m};N)\)
        \STATE Restore residualised DC coefficients and denormalise with \(\Phi\)
        \STATE \(\hat{\mathbf{Y}}_j \gets \Phi^{-1}(\hat{\tilde{\mathbf{Z}}}_{0,j})\) \COMMENT{Inverse DCT to time-domain representation}
        \STATE \(\hat{\mathbf{X}}_j \gets R^{-1}(\hat{\mathbf{Y}}_j,\mathbf{C}_j)\) \COMMENT{For displacement models, add the conditioning structure}
        \IF{coordinate refiner or SHAKE projection is enabled}
            \STATE Refine \(\hat{\mathbf{X}}_j\) and project adjacent \(C_\alpha\)--\(C_\alpha\) distances toward \(3.8\,\text{\AA}\)
        \ENDIF
        \STATE Append valid residues of \(\hat{\mathbf{X}}_j\) to each output trajectory
        \IF{chained generation is enabled}
            \STATE \(\mathbf{C}^{(b)}_{j+1} \gets \hat{\mathbf{X}}^{(b)}_{j,T}\) for all \(b\)
        \ELSE
            \STATE \(\mathbf{C}^{(b)}_{j+1} \gets \mathbf{X}^{(b)}_{\mathrm{nat}}\) for all \(b\)
        \ENDIF
    \ENDFOR
    \STATE Concatenate windows, trim to \(F\) frames, optionally minimise the full trajectory, and export PDB/XTC files
    \STATE \textbf{return} \(\{\hat{\mathbf{X}}^{(b)}_{1:F}\}_{b=1}^B\)
  \end{algorithmic}
\end{algorithm}

\subsection{\(C_\alpha\) Energy Minimisation}
\label{sec:appendix_energy_min}

As an optional post-processing step, generated trajectories can be relaxed with a lightweight \(C_\alpha\)-only energy minimiser. This is not intended to be a full physical force field. Instead, it is a geometry-cleanup objective that reduces nonlocal \(C_\alpha\)--\(C_\alpha\) clashes while preserving the local shape of the generated trajectory.

For each generated trajectory, coordinates are flattened over frames and optimised in batches of 50 structures. We construct a fixed \(C_\alpha\) topology containing adjacent bonds \((i,i+1)\), bend angles \((i,i+1,i+2)\), pseudo-dihedrals \((i,i+1,i+2,i+3)\), and nonbonded \(C_\alpha\)--\(C_\alpha\) pairs with sequence separation at least two. Nonbonded pairs are cached every 10 optimisation steps by retaining pairs that are within \(10\,\text{\AA}\) in any structure in the current batch.

The minimised energy is
\begin{align}
E(\mathbf{X}) &=
k_b \sum_{(i,j)\in\mathcal{B}} m_{ij}
  \left(d_{ij}-d^{0}_{ij}\right)^2
+ k_\theta \sum_{(i,j,k)\in\mathcal{A}} m_{ijk}
  \Delta(\theta_{ijk},\theta^{0}_{ijk})^2 \nonumber \\
&\quad
+ k_\phi \sum_{(i,j,k,l)\in\mathcal{D}} m_{ijkl}
  \Delta(\phi_{ijkl},\phi^{0}_{ijkl})^2
+ k_{\mathrm{nb}} \sum_{(i,j)\in\mathcal{N}} m_{ij}
  \left[r_{\mathrm{nb}}-d_{ij}\right]_+^2 ,
\end{align}
where \(d_{ij}\) is a \(C_\alpha\)--\(C_\alpha\) distance, \(\theta\) and \(\phi\) are the \(C_\alpha\) bend-angle and pseudo-dihedral angle, \(\Delta(a,b)=\mathrm{atan2}(\sin(a-b),\cos(a-b))\) handles angle periodicity, and \(m\) masks invalid residues. Angle and dihedral targets are set from the generated input coordinates, so minimisation is biased toward preserving the sampled conformation. For adjacent bonds, the target \(d^0_{ij}\) is the initial generated bond length if it already lies in \([3.57,4.11]\,\text{\AA}\), and otherwise the ideal \(3.8\,\text{\AA}\). In the mdCATH protocol we use \(k_b=10000\), \(k_\theta=100\), \(k_\phi=10\), \(k_{\mathrm{nb}}=250\), and \(r_{\mathrm{nb}}=3.5\,\text{\AA}\). An optional segment--segment self-intersection penalty is implemented but disabled in the reported default settings.

The default protocol first runs a short Adam warm-start stage for up to 50 steps with \(k_{\mathrm{nb}}=100\), then an L-BFGS stage for 30 outer steps with 10 inner iterations per step. Optimisation stops early when the average number of cached nonbonded clashes below \(3.5\,\text{\AA}\) is at most 0.7 per structure. During inference the minimiser can be applied either to each generated window independently or to the concatenated trajectory after all windows have been stacked.

\subsection{Reconstruction Error}
\label{sec:appendix_reconstruction_error}
Table~\ref{tab:dct_dft_truncation_errors} details the DCT and DFT reconstruction errors for different spectral volume truncation across a number of metrics over the whole train set.

\begin{table*}[!b]
  \caption{DCT and DFT truncation reconstruction errors for native-frame C$_\alpha$ displacement at matched fractions of the usable spectrum. Values are mean $\pm$ standard deviation over the same sampled trajectories used for the reconstruction-error figure. $\uparrow$: higher is better; $\downarrow$: lower is better.}
  \label{tab:dct_dft_truncation_errors}
  \begin{center}
    \begin{small}
      \begin{sc}
      \setlength{\tabcolsep}{3pt}
        \begin{tabular}{llcccccc}
        \toprule
         & Spectrum & CA RMSD (\AA) $\downarrow$ & Boundary RMSD (\AA) $\downarrow$ & C$_\alpha$--C$_\alpha$ W$_1$ (\AA) $\downarrow$ & RMSF $\rho$ $\uparrow$ & Energy $\uparrow$ \\
        \midrule
        DCT & 6.2\% & 1.934 $\pm$ 1.620 & 1.841 $\pm$ 1.546 & 0.480 & 0.976 & 0.837 \\
        DCT & 12.5\% & 1.650 $\pm$ 1.371 & 1.541 $\pm$ 1.292 & 0.374 & 0.987 & 0.877 \\
        DCT & 25.0\% & 1.353 $\pm$ 1.102 & 1.288 $\pm$ 1.064 & 0.270 & 0.994 & 0.914 \\
        DCT & 50.0\% & 0.997 $\pm$ 0.791 & 0.918 $\pm$ 0.746 & 0.173 & 0.998 & 0.951 \\
        DCT & 75.0\% & 0.679 $\pm$ 0.534 & 0.573 $\pm$ 0.468 & 0.123 & 0.999 & 0.977 \\
        DCT & 100.0\% & 0.000 $\pm$ 0.000 & 0.000 $\pm$ 0.000 & 0.000 & 1.000 & 1.000 \\
        DFT & 6.2\% & 1.986 $\pm$ 1.650 & 2.543 $\pm$ 1.956 & 0.496 & 0.975 & 0.889 \\
        DFT & 12.5\% & 1.680 $\pm$ 1.389 & 2.043 $\pm$ 1.594 & 0.383 & 0.987 & 0.917 \\
        DFT & 25.0\% & 1.371 $\pm$ 1.113 & 1.599 $\pm$ 1.250 & 0.275 & 0.993 & 0.943 \\
        DFT & 50.0\% & 1.009 $\pm$ 0.799 & 1.148 $\pm$ 0.890 & 0.176 & 0.998 & 0.968 \\
        DFT & 75.0\% & 0.680 $\pm$ 0.533 & 0.770 $\pm$ 0.599 & 0.124 & 0.999 & 0.985 \\
        DFT & 100.0\% & 0.000 $\pm$ 0.000 & 0.000 $\pm$ 0.000 & 0.000 & 1.000 & 1.000 \\
        \bottomrule
        \end{tabular}
      \end{sc}
    \end{small}
  \end{center}
  \vskip -0.1in
\end{table*}

\subsection{Evaluation}
\label{sec:appendix_evaluation}

\paragraph{Benchmark datasets and trajectory construction.}
We evaluate on two test sets, the mdCATH held-out test set which uses the same train/val/test split as MarS-FM~\cite{jingAlphaFoldMeetsFlow2024c,kapusniakMarSFMGenerativeModeling2026}, and the 82 domain ATLAS test set. For mdCATH, each domain is simulated with five replicas at each of five temperatures
\(T_\text{temp} \in \{320,348,379,413,450\}\,\mathrm{K}\), sampled at \(1\,\mathrm{ns}\) per frame. For ATLAS, each domain has three trajectories at \(300\,\mathrm{K}\); we subsample each trajectory with stride \(100\) so that the evaluation resolution matches the \(1\,\mathrm{ns}\) timestep used by our model. Following the ensemble evaluation protocol used by MarS-FM~\cite{kapusniakMarSFMGenerativeModeling2026}, we generate multiple trajectories and compare to pooled MD ensembles for each target, followed by averaging over the results.

For mdCATH, one generated sample is formed by predicting two 256-frame windows conditioned at window positions \(s=0\) and \(s=256\), concatenating them, and trimming down to a 500-frame trajectory. For ATLAS, one generated sample is a single 256-frame trimmed to the first 100 frames. We generate \(5\) independent samples per target on mdCATH and \(3\) independent samples per target on ATLAS, evaluate each sample against the pooled MD ensemble, and average the resulting per-repeat metrics.

\paragraph{Alignment and oracle baseline.}
All generated and reference trajectories are rigid-body aligned to the first frame of the first MD replica for that target. We also report an \emph{oracle} MD upper bound via leave-one-out pooling: for mdCATH, one replica is held out and compared against the pool of the remaining four, repeated over all five choices; for ATLAS, one replica is held out and compared against the remaining two. The oracle therefore measures intrinsic replicate-to-replicate agreement under the same metrics, providing a realistic ceiling for what any stochastic emulator can achieve.

\paragraph{RMSF and correlation metrics.}
For a trajectory \(\mathbf{X} \in \mathbb{R}^{T \times L \times 3}\), the per-residue root-mean-square fluctuation (RMSF) is
\begin{equation}
  \mathrm{RMSF}_i(\mathbf{X}) =
  \sqrt{\frac{1}{T}\sum_{t=1}^{T}\left\|\mathbf{X}_{t,i,:} -
  \bar{\mathbf{X}}_{i,:}\right\|_2^2},
  \qquad
  \bar{\mathbf{X}}_{i,:}=\frac{1}{T}\sum_{t=1}^{T}\mathbf{X}_{t,i,:}.
\end{equation}
RMSF is a direct measure of residue-wise flexibility, so it is especially important for testing how well our model captures conformational dynamics especially at high temperatures. For each target \(n\), we compute a per-target Pearson correlation
\begin{equation}
  r_{\mathrm{RMSF}}^{(n)} =
  \mathrm{corr}\!\left(
  \mathrm{RMSF}(\hat{\mathbf{X}}^{(n)}),
  \mathrm{RMSF}(\mathbf{X}_{\mathrm{MD}}^{(n)})
  \right),
\end{equation}
and report the median across targets. We also compute the related per-target Spearman correlation \(\rho_{\mathrm{RMSF}}^{(n)}\), which is less sensitive to amplitude calibration and instead emphasizes correct flexibility ranking - we prioritise Pearson as we observed the model struggles more in matching magnitude than residue profiles and this is the metric most other models report~\cite{kapusniakMarSFMGenerativeModeling2026}. Global RMSF correlation pools all residues from all targets into one concatenated vector before computing a single Pearson \(r\) or Spearman \(\rho\). Thus, global RMSF tests overall flexibility calibration across the entire benchmark, whereas per-target RMSF tests whether the model gets the residue-level fluctuation profile right for each individual protein.

\paragraph{Pairwise RMSD correlation and distribution.}
For each target, we estimate the mean pairwise C\(_\alpha\)-RMSD within an ensemble,
\begin{equation}
  \overline{\mathrm{RMSD}}(\mathbf{X}) =
  \frac{2}{T(T-1)}\sum_{1 \leq t < t' \leq T}
  \mathrm{RMSD}\!\left(\mathbf{X}_{t}, \mathbf{X}_{t'}\right),
\end{equation}
and report the Pearson correlation across targets between the generated and MD values of \(\overline{\mathrm{RMSD}}\) - evaluating how the model reproduces structural diversity for each protein. In addition, we compare the full distribution of sampled pairwise RMSD values between predicted and MD ensembles using the Jensen--Shannon divergence (JSD). Whereas the correlation metric tests calibration of the target-level summary statistic, the JSD tests whether the overall diversity distribution is reproduced.

\paragraph{Binned distributional metrics.}
Several scalar observables are compared via histogram-based divergences. Given scalar samples \(u\), we estimate discrete distributions \(P\) and \(Q\) using 100-bin histograms over a shared range, add a floor \(\epsilon = 10^{-5}\), and renormalise. The forward Kullback--Leibler (KL) divergence is
\begin{equation}
  D_{\mathrm{KL}}(P\|Q) = \sum_b P_b \log \frac{P_b}{Q_b},
\end{equation}
and the Jensen--Shannon divergence is
\begin{equation}
  \mathrm{JSD}(P,Q)
  = \frac{1}{2} D_{\mathrm{KL}}(P\|M)
  + \frac{1}{2} D_{\mathrm{K§L}}(Q\|M),
  \qquad
  M = \frac{1}{2}(P+Q).
\end{equation}
These distributional comparisons are used for pairwise RMSD, RMSF, radius of gyration, fraction of native contacts, Global Distance Test -- Total Score (GDT-TS), and spectral amplitudes.

\paragraph{RMWD and PCA-based ensemble distances.}
To compare full positional ensembles, we approximate the generated and reference positional distributions for atom \(i\) by Gaussians \(\mathcal{N}(\boldsymbol{\mu}^{\mathrm{pred}}_i,\Sigma^{\mathrm{pred}}_i)\) and \(\mathcal{N}(\boldsymbol{\mu}^{\mathrm{gt}}_i,\Sigma^{\mathrm{gt}}_i)\). The squared 2-Wasserstein distance is
\begin{equation}
  \mathcal{W}_2^2(i) =
  \|\boldsymbol{\mu}^{\mathrm{pred}}_i - \boldsymbol{\mu}^{\mathrm{gt}}_i\|_2^2
  + \mathrm{Tr}\!\left(
    \Sigma^{\mathrm{pred}}_i + \Sigma^{\mathrm{gt}}_i
    - 2\Big[
      (\Sigma^{\mathrm{gt}}_i)^{1/2}
      \Sigma^{\mathrm{pred}}_i
      (\Sigma^{\mathrm{gt}}_i)^{1/2}
    \Big]^{1/2}
  \right).
\end{equation}
The root mean Wasserstein distance (RMWD) is then
\begin{equation}
  \mathrm{RMWD}
  =
  \sqrt{\frac{1}{L}\sum_{i=1}^{L}\mathcal{W}_2^2(i)}.
\end{equation}
RMWD penalises both mean positional shifts and covariance mismatch, so it is more sensitive than RMSF to errors in the full spatial distribution of each residue.

We also compute PCA-based ensemble distances. PCA is fit on the reference MD ensemble, both predicted and reference structures are projected onto the first two reference PCs, projected distances are normalised by \(\sqrt{L}\), and an assignment-based empirical 2-Wasserstein distance is computed between equal-sized samples. This asks whether the generated ensemble covers the same dominant collective modes as MD.

\paragraph{tICA free-energy visualisations.}
For the case-study visualisations in Figure~\ref{fig:results}, we additionally project trajectories onto slow collective coordinates using tICA. For each domain and temperature, we construct native-displacement C\(_\alpha\) features by flattening \(\mathbf{X}_t-\mathbf{X}_{\mathrm{nat}}\) over residues and coordinates, and fit a two-component tICA basis jointly on the aligned MD and generated trajectories using a lag time of $\tau=10$ frames. Both trajectories are projected onto the first two tICs. We then estimate a shared two-dimensional histogram density \(p(y_1,y_2)\) and plot the dimensionless relative free-energy surface
\begin{equation}
  F(y_1,y_2) = -\log p(y_1,y_2) + C,
\end{equation}
where \(C\) shifts the minimum finite value to zero. These overlays give a visual gauge of how generated trajectories occupy the same slow-mode basins as the MD reference.

It is worth noting, these plots a largely qualitative reporter of ensemble overlap. Smaller proteins tend to show more overlap in our tICA space overlays (results not shown) likely due to the increased dimensionality of the conformational landscape in larger proteins and the use of just the top two components.

\paragraph{Principal-component similarity and contact metrics.}
Let \(\mathbf{v}^{\mathrm{pred}}_1\) and \(\mathbf{v}^{\mathrm{gt}}_1\) denote the leading principal components of the generated and reference ensembles. We compute the unsigned cosine similarity
\begin{equation}
  s_{\mathrm{PC1}} =
  \left|
  \frac{\langle \mathbf{v}^{\mathrm{pred}}_1, \mathbf{v}^{\mathrm{gt}}_1 \rangle}
       {\|\mathbf{v}^{\mathrm{pred}}_1\|_2 \, \|\mathbf{v}^{\mathrm{gt}}_1\|_2}
  \right|,
\end{equation}
and summarise the percentage of targets with \(s_{\mathrm{PC1}} > 0.5\). This metric tests whether the model recovers the direction of the dominant collective motion.

For contact metrics, we form C\(_\alpha\) contact probabilities using an 8\,\AA{} threshold. \emph{Weak contacts} are native contacts whose contact probability drops below \(0.9\), and \emph{transient contacts} are non-native contacts whose contact probability rises above \(0.1\). We compare the predicted and MD weak/transient contact sets using the Jaccard similarity
\begin{equation}
  J(A,B)=\frac{|A\cap B|}{|A\cup B|}.
\end{equation}
These metrics test whether the model captures the same topological diversity as MD, often more mechanistically meaningful than raw coordinate error.

\paragraph{Native-contact, folding, and structural-validity metrics.}
From the native structure we extract the set of non-local native C\(_\alpha\) contacts and compute the fraction of native contacts retained in each frame using the smooth logistic definition from prior folding-emulation work. This yields a per-frame Fraction of Native Contacts trajectory \(\mathrm{FNC}(t)\). We compare predicted and MD FNC distributions using JSD, and also report the mean predicted and reference FNC.

Using a folding threshold \(\mathrm{FNC}>0.5\), we estimate the folded-state probability \(p_{\mathrm{fold}}\) and corresponding folding free energy
\begin{equation}
  \Delta G_{\mathrm{fold}}
  = -kT \log \frac{p_{\mathrm{fold}}}{1-p_{\mathrm{fold}}},
\end{equation}
where \(kT\) uses the evaluation temperature. We report predicted and reference \(\Delta G_{\mathrm{fold}}\), the absolute folding free-energy error, and the corresponding folded fractions. These quantities test whether the generated ensemble preserves the correct balance between folded and unfolded states.

We additionally compute per-frame GDT-TS to the native structure, yielding both mean GDT-TS and a GDT-TS JSD. This measures native-state structural similarity at multiple distance cutoffs and is useful for checking whether the model preserves native-like geometry while still producing non-trivial motion. Finally, we report mean consecutive C\(_\alpha\) distances and their JSD, serving as a lightweight geometric-validity diagnostic analogous to a bond-length sanity check.

\paragraph{Radius of gyration and compactness.}
For each frame we compute the C\(_\alpha\)-based radius of gyration
\begin{equation}
  R_g(\mathbf{X}_t) =
  \sqrt{\frac{1}{L}\sum_{i=1}^{L}
  \left\|\mathbf{X}_{t,i,:} - \bar{\mathbf{X}}_{t,:}\right\|_2^2 }.
\end{equation}
We compare predicted and MD \(R_g\) distributions using both JSD and forward KL, and also report the mean predicted and reference values as a measure of global compactness consistency.

\paragraph{Spectral metrics.}
For our spectral model, we evaluate the predicted spectral volume directly. Let \(\hat{\mathbf{Z}}\) and \(\mathbf{Z}\) denote predicted and ground-truth normalised spectral tensors. We compute the spectral mean-squared error
\begin{equation}
  \mathrm{SpecMSE}
  =
  \frac{1}{LKC}
  \sum_{l=1}^{L}\sum_{k=0}^{K-1}\sum_{c=1}^{C}
  \left(\hat{Z}_{l,k,c} - Z_{l,k,c}\right)^2.
\end{equation}
We also compute the per-residue amplitude at frequency \(k\),
\begin{equation}
  A_{l,k}(\mathbf{Z}) = \|\mathbf{Z}_{l,k,:}\|_2,
\end{equation}
and aggregate amplitudes into DC (\(k=0\)), low-, mid-, high-, and total-frequency bands. From these amplitudes we report bandwise Pearson correlations, JSDs, and amplitude-recovery scores. For a reference amplitude vector \(a^{\mathrm{ref}}\) and predicted amplitude vector \(a^{\mathrm{pred}}\), the amplitude-recovery score is
\begin{equation}
  \mathrm{Rec}(a^{\mathrm{ref}},a^{\mathrm{pred}})
  =
  \frac{1}{N}\sum_{i=1}^{N}
  \left(
  1 -
  \frac{|a_i^{\mathrm{pred}} - a_i^{\mathrm{ref}}|}
       {a_i^{\mathrm{pred}} + a_i^{\mathrm{ref}} + \varepsilon}
  \right),
\end{equation}
with \(\varepsilon=10^{-8}\). These diagnostics are especially important because RMSF magnitude is dominated by the DC and neighboring low-frequency modes. For example, the model can achieve reasonable coordinate MSE while still underestimating the low-\(k\) amplitudes that control large-scale flexibility. We therefore report DC-specific metrics - DC Pearson correlation, DC JSD, and DC amplitude recovery - alongside broader low/mid/high/total band metrics.

\begin{figure*}[!t]
  \centering
  \includegraphics[width=\textwidth]{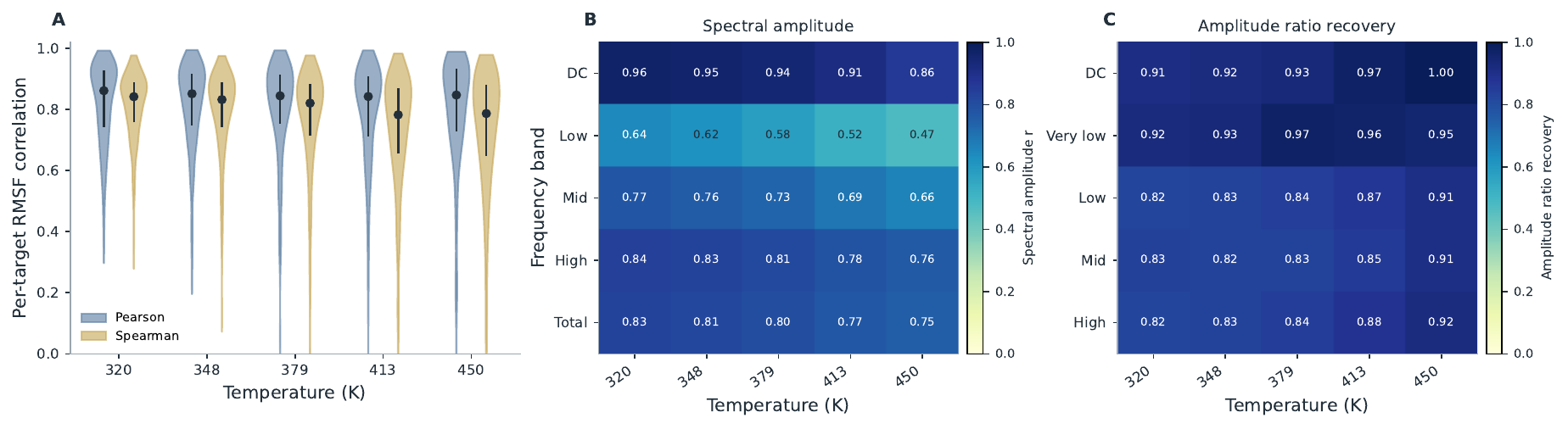}
  \caption{Residue flexibility (RMSF) and spectral volumes prediction accuracy across temperatures on the mdCATH test set. \textbf{A} Violin plots show the distribution of RMSF Pearson and Spearman correlations (predicted vs reference MD trajectory following the protocol in Appendix~\ref{sec:appendix_evaluation} across temperatures. We also show spectral volume prediction accuracy across different regions of the spectral volume defined by grouping frequencies according to Table~\ref{tab:freq_groups} at different temperatures. Accuracy is assessed by comparing predicted 256 frame window spectral volumes against same starting timepoint 256 window for MD trajectory references averaged across 5 repeats. \textbf{B} Pearson r correlation. \textbf{C} Ratio of summed spectral amplitudes of predicted vs reference.}
  \label{fig:spectral_results}
\end{figure*}

\begin{table}[t]
  \caption{Frequency bands used for the frequency-resolved spectral-volume accuracy analysis in Figure~\ref{fig:results}. Bands are defined over DCT mode index \(k\) for 256-frame trajectory windows. The grouping separates the trajectory mean, slow collective drift, low-frequency conformational changes, intermediate backbone-scale motion, and fast local fluctuations. Timescales are approximate because DCT basis functions are finite-window cosine modes rather than periodic Fourier modes.}
  \label{tab:freq_groups}
  \begin{center}
    \begin{small}
      \begin{sc}
      \setlength{\tabcolsep}{5pt}
        \begin{tabular}{lccc}
        \toprule
        Band & DCT modes \(k\) & Approx.\ timescale \\ 
        \midrule
        DC & \(0\) & trajectory mean \\ 
        Very low & \(1\)--\(4\) & \(\gtrsim 128\,\mathrm{ns}\) \\ 
        Low & \(5\)--\(16\) & \(32\)--\(102\,\mathrm{ns}\) \\ 
        Mid & \(17\)--\(64\) & \(8\)--\(30\,\mathrm{ns}\) \\ 
        High & \(65\)--\(128\) & \(\lesssim 8\,\mathrm{ns}\) \\ 
        \bottomrule
        \end{tabular}
      \end{sc}
    \end{small}
  \end{center}
  \vskip -0.1in
\end{table}

\subsection{Spectral Volume Prediction Accuracy}
\label{sec:appendix_spectral_volume_pred}
Figure~\ref{fig:spectral_results}B \& C shows spectral prediction accuracy measurements across temperatures and frequency groupings (Table~\ref{tab:freq_groups}). 
The high frequencies are clearly the easiest to predict (Figure~\ref{fig:results}), which is consistent with the effectiveness of diffusion models in the white-noise regime. In contrast, we observed that the low frequencies, characterised by a highly structured signal, remained the most difficult to predict.

This frequency-resolved error attribution motivated the dedicated low-$k$ amplitude-calibration pathway (Section~\ref{sec:appendix_architectures}) which we found improved DC component prediction substantially, as shown by the high correlation in Figure~\ref{fig:spectral_results}B. Directly improving mean displacement prediction like so is not available to time-domain methods, where reconstruction errors are entangled across all temporal scales.

\begin{table*}[!t]
  \caption{Trajectory benchmark on mdCATH at 320\,K ($n=478$ test targets). BioEMU and MarS-FM results from the MarS-FM paper (mean $\pm$ SEM). Oracle uses the held-out MD trajectory as prediction. For DynaMode, $\pm$ denotes bootstrap SEM across targets. $\uparrow$: higher is better; $\downarrow$: lower is better. \textbf{Bold}: best per row among non-oracle methods.}
  \label{tab:mdcath_trajectory_mars_320k}
  \begin{center}
    \begin{small}
      \begin{sc}
      \setlength{\tabcolsep}{4pt}
        \begin{tabular}{lcccc}
        \toprule
        Metric & Oracle (320\,K) & BioEMU & MarS-FM & DynaMode (320\,K) \\
        \midrule
        Pair.\ RMSD $r$ $\uparrow$ & 0.984 & $0.580\pm.001$ & $\mathbf{0.900\pm.001}$ & $0.833\pm.028$ \\
        Global RMSF $r$ $\uparrow$ & 0.894 & $0.630\pm.004$ & $\mathbf{0.870\pm.001}$ & $0.823\pm.017$ \\
        Per-tgt RMSF $r$ $\uparrow$ & 0.894 & $0.840\pm.002$ & $\mathbf{0.900\pm.003}$ & $0.861\pm.007$ \\
        RMWD $\downarrow$ & 1.85 & -- & -- & $2.73\pm.102$ \\
        PCA $\mathcal{W}_2$ $\downarrow$ & 1.33 & -- & -- & $1.63\pm.070$ \\
        Weak J $\uparrow$ & 0.780 & -- & -- & $0.538\pm.004$ \\
        Trans.\ J $\uparrow$ & 0.508 & -- & -- & $0.258\pm.003$ \\
        $R_g$ JSD $\downarrow$ & 0.09 & $0.36\pm.001$ & $\mathbf{0.14\pm.001}$ & $0.15\pm.004$ \\
        $\Delta G$ MAE $\downarrow$ & 0.60 & $0.83\pm.003$ & $\mathbf{0.58\pm.001}$ & $1.23\pm.119$ \\
        \bottomrule
        \end{tabular}
      \end{sc}
    \end{small}
  \end{center}
  \vskip -0.1in
\end{table*}

\begin{table*}[!t]
  \caption{Trajectory benchmark on mdCATH at 450\,K ($n=487$ test targets). BioEMU and MarS-FM results from the MarS-FM paper (mean $\pm$ SEM). Oracle uses the held-out MD trajectory as prediction. For DynaMode, $\pm$ denotes bootstrap SEM across targets. $\uparrow$: higher is better; $\downarrow$: lower is better. \textbf{Bold}: best per row among non-oracle methods.}
  \label{tab:mdcath_trajectory_mars_450k}
  \begin{center}
    \begin{small}
      \begin{sc}
      \setlength{\tabcolsep}{4pt}
        \begin{tabular}{lcccc}
        \toprule
        Metric & Oracle (450\,K) & BioEMU & MarS-FM & DynaMode (450\,K) \\
        \midrule
        Pair.\ RMSD $r$ $\uparrow$ & 0.990 & $0.250\pm.002$ & $\mathbf{0.650\pm.004}$ & $0.590\pm.034$ \\
        Global RMSF $r$ $\uparrow$ & 0.862 & $0.410\pm.004$ & $\mathbf{0.710\pm.003}$ & $0.691\pm.022$ \\
        Per-tgt RMSF $r$ $\uparrow$ & 0.862 & $0.660\pm.001$ & $\mathbf{0.890\pm.001}$ & $0.847\pm.009$ \\
        RMWD $\downarrow$ & 4.82 & -- & -- & $6.38\pm.100$ \\
        PCA $\mathcal{W}_2$ $\downarrow$ & 3.37 & -- & -- & $4.04\pm.080$ \\
        Weak J $\uparrow$ & 0.930 & -- & -- & $0.715\pm.003$ \\
        Trans.\ J $\uparrow$ & 0.511 & -- & -- & $0.191\pm.004$ \\
        $R_g$ JSD $\downarrow$ & 0.07 & $0.40\pm.001$ & $\mathbf{0.10\pm.001}$ & $0.12\pm.003$ \\
        $\Delta G$ MAE $\downarrow$ & 1.58 & $4.67\pm.004$ & $\mathbf{1.05\pm.003}$ & $2.50\pm.111$ \\
        \bottomrule
        \end{tabular}
      \end{sc}
    \end{small}
  \end{center}
  \vskip -0.1in
\end{table*}

\subsection{Temperature Stratified Evaluation Against MarS-FM}
\label{sec:appendix_temp_mdcath_eval}
For direct comparison against the recent MarS-FM we detail our trajectory benchmark results on the mdCATH test set limited to 320K (Table~\ref{tab:mdcath_trajectory_mars_320k}) and 450K (Table~\ref{tab:mdcath_trajectory_mars_450k}).

\subsection{Structural Validity}
\label{sec:appendix_structural_validity}

\paragraph{Nonbonded and backbone-trace distances.}
To diagnose steric collapse in C\(_\alpha\)-only trajectories, we compute nonbonded C\(_\alpha\)--C\(_\alpha\) distances for all residue pairs separated by at least two positions in sequence,
\begin{equation}
  d_{ij}(t) = \left\|\mathbf{X}_{t,i}-\mathbf{X}_{t,j}\right\|_2,
  \qquad |i-j|\geq 2.
\end{equation}
We report the per-frame minimum distance and the number or fraction of frames with at least one pair below 3.5, 3.0, or 2.5\,\AA{}. For the nonlocal backbone-trace distance shown in Figure~\ref{fig:results}, each consecutive C\(_\alpha\) pair defines a line segment \(S_i(t)=[\mathbf{X}_{t,i},\mathbf{X}_{t,i+1}]\). For segment pairs with \(|i-j|\geq 2\), we compute
\begin{equation}
  d^{\mathrm{seg}}_{ij}(t) =
  \min_{u,v\in[0,1]}
  \left\|
  (1-u)\mathbf{X}_{t,i}+u\mathbf{X}_{t,i+1}
  -
  (1-v)\mathbf{X}_{t,j}+v\mathbf{X}_{t,j+1}
  \right\|_2 .
\end{equation}
Segment distances below 1.0\,\AA{} indicate severe trace contacts, and distances below 0.25\,\AA{} are treated as trace-intersection proxies.

Structural validity metrics across the mdCATH and ATLAS test sets are collected in Table~\ref{tab:v12a_structural_validity_bond_normalised} including neighbouring C$_\alpha$--C$_\alpha$ distance statistics, FNC, GDT-TS, LDDT, and the average minimum of nonlocal (non-neighbouring) C$_\alpha$--C$_\alpha$ distances and the fraction of structures in each trajectory with at least one C$_\alpha$--C$_\alpha < 1.0\AA$ as reporters of steric clashes.

Although C$_\alpha$--C$_\alpha$ distances, FNC, GDT-TS and LDDT show good structural integrity compared to the MD references, there is a significant number of steric clashes reported. Whilst generally we can expect more steric clashes from any generative model than the reference MD, 20 and 6.56 of nonbonded C$_\alpha$--C$_\alpha$ on average per frame showing steric clashes, and 91.8\% and 87.6\% of frames per trajectory with at least one steric clash for mdCATH and ATLAS respectively is likely a direct result of our DCT representation. These results are the raw model inference output, not post-inference energy minimised. Particularly concerning is backbone trace nonlocal C$_\alpha$--C$_\alpha$ segments $<1.0\dot{A}$ occurring at least once in 51\% and 18.4\% of mdCATH and ATLAS test set generated structures respectively which are possible grounds for topological changes from chain crossover which would not be fixable by energy minimsation.

\begin{table*}[!t]
  \caption{Structural validity metrics split by dataset and temperature. mdCATH overall and per-temperature columns are followed by a vertically separated ATLAS 300 K column; each condition reports the MD reference and predicted ensemble side by side.}
  \label{tab:v12a_structural_validity_bond_normalised}
  \begin{center}
    \begin{scriptsize}
      \setlength{\tabcolsep}{2.6pt}
      \resizebox{\textwidth}{!}{
        \begin{tabular}{lcccccccccccc|cc}
        \toprule
        Measure & \multicolumn{2}{c}{mdCATH all} & \multicolumn{2}{c}{mdCATH 320 K} & \multicolumn{2}{c}{mdCATH 348 K} & \multicolumn{2}{c}{mdCATH 379 K} & \multicolumn{2}{c}{mdCATH 413 K} & \multicolumn{2}{c}{mdCATH 450 K} & \multicolumn{2}{c}{ATLAS 300 K} \\
        \cmidrule(lr){2-3} \cmidrule(lr){4-5} \cmidrule(lr){6-7} \cmidrule(lr){8-9} \cmidrule(lr){10-11} \cmidrule(lr){12-13} \cmidrule(lr){14-15}
        & MD & Pred & MD & Pred & MD & Pred & MD & Pred & MD & Pred & MD & Pred & MD & Pred \\
        \midrule
        C$_\alpha$--C$_\alpha$ mean (\AA) & 3.831 & 3.804 & 3.833 & 3.808 & 3.832 & 3.805 & 3.831 & 3.803 & 3.830 & 3.802 & 3.829 & 3.804 & 3.836 & 3.813 \\
        C$_\alpha$--C$_\alpha$ std (\AA) & 0.009 & 0.009 & 0.008 & 0.009 & 0.008 & 0.009 & 0.008 & 0.009 & 0.009 & 0.009 & 0.010 & 0.009 & 0.006 & 0.007 \\
        C$_\alpha$--C$_\alpha$ 1st--99th pct. width (\AA) & 0.041 & 0.044 & 0.038 & 0.045 & 0.040 & 0.043 & 0.040 & 0.042 & 0.041 & 0.042 & 0.043 & 0.044 & 0.031 & 0.033 \\
        FNC mean & 0.613 & 0.696 & 0.774 & 0.848 & 0.731 & 0.804 & 0.658 & 0.738 & 0.541 & 0.622 & 0.360 & 0.464 & 0.899 & 0.907 \\
        GDT-TS mean & 0.447 & 0.555 & 0.601 & 0.728 & 0.555 & 0.674 & 0.479 & 0.594 & 0.375 & 0.469 & 0.226 & 0.308 & 0.656 & 0.764 \\
        LDDT mean & 0.617 & 0.654 & 0.737 & 0.794 & 0.704 & 0.751 & 0.649 & 0.690 & 0.563 & 0.585 & 0.431 & 0.447 & 0.837 & 0.839 \\
        Mean min nonlocal C$_\alpha$--C$_\alpha$ (\AA) & 3.488 & 1.339 & 3.513 & 2.206 & 3.504 & 1.835 & 3.497 & 1.352 & 3.480 & 0.858 & 3.446 & 0.439 & 3.443 & 2.188 \\
        Frames with any nonbonded C$_\alpha$--C$_\alpha$ $<$ 3.5 \AA (\%) & 0.4 & 91.8 & 0.3 & 80.2 & 0.3 & 87.6 & 0.4 & 93.0 & 0.4 & 98.2 & 0.5 & 99.9 & 1.3 & 87.6 \\
        Frames with any nonbonded C$_\alpha$--C$_\alpha$ $<$ 3.0 \AA (\%) & 0.0 & 84.2 & 0.0 & 65.6 & 0.0 & 76.1 & 0.0 & 85.1 & 0.0 & 94.8 & 0.0 & 99.3 & 0.0 & 72.1 \\
        Nonbonded C$_\alpha$--C$_\alpha$ pair clashes/frame $<$ 3.5 \AA & 0.005 & 20.755 & 0.003 & 5.909 & 0.003 & 8.485 & 0.004 & 13.372 & 0.007 & 25.742 & 0.008 & 49.996 & 0.013 & 6.566 \\
        Nonbonded C$_\alpha$--C$_\alpha$ pair clashes/frame $<$ 3.0 \AA & 0.001 & 12.761 & 0.000 & 3.339 & 0.000 & 4.911 & 0.000 & 7.958 & 0.002 & 15.798 & 0.002 & 31.624 & 0.000 & 3.394 \\
        Nonbonded C$_\alpha$--C$_\alpha$ pair clashes/frame $<$ 2.5 \AA & 0.001 & 7.320 & 0.000 & 1.819 & 0.000 & 2.713 & 0.000 & 4.468 & 0.002 & 9.037 & 0.002 & 18.458 & 0.000 & 1.762 \\
        Nonbonded C$_\alpha$--C$_\alpha$ pair clashes/clashing frame $<$ 2.5 \AA & 87.488 & 9.659 & 0.000 & 3.519 & 0.000 & 4.271 & 0.000 & 5.923 & 105.312 & 10.067 & 76.080 & 18.831 & 0.000 & 3.201 \\
        Nonbonded C$_\alpha$--C$_\alpha$ pair clashes/trajectory $<$ 3.5 \AA & 11.4 & 51888.6 & 5.9 & 14772.3 & 7.4 & 21213.4 & 8.4 & 33429.8 & 16.5 & 64354.0 & 18.9 & 124990.0 & 3.8 & 3283.2 \\
        Segment crossings/frame $<$ 1.0 \AA & 0.001 & 4.146 & 0.000 & 0.791 & 0.000 & 1.301 & 0.000 & 2.571 & 0.002 & 5.268 & 0.002 & 10.854 & 0.000 & 0.621 \\
        Frames with nonlocal C$_\alpha$--C$_\alpha$ segment $<$ 1.0 \AA (\%) & 0.0 & 51.0 & 0.0 & 21.2 & 0.0 & 31.0 & 0.0 & 47.8 & 0.0 & 68.3 & 0.0 & 87.1 & 0.0 & 18.4 \\
        \bottomrule
        \end{tabular}
      }
    \end{scriptsize}
  \end{center}
  \vskip -0.1in
\end{table*}

\end{document}